\documentclass[%
 reprint,
 superscriptaddress,
 nofootinbib,
 amsmath,amssymb,
 aps,prd
]{revtex4-2}

\usepackage[dvipsnames]{xcolor}
\usepackage{graphicx}
\usepackage{dcolumn}
\usepackage{bm}
\usepackage{placeins}
\usepackage{orcidlink}
\usepackage{float}
\usepackage{comment}
\usepackage{graphicx}
\usepackage[normalem]{ulem}

\usepackage{tabularx}
\usepackage{adjustbox}
\usepackage{fancyhdr} 

\usepackage{hyperref}
\usepackage{makecell}

\usepackage[caption=false]{subfig}
\usepackage{hhline}

\newcommand{\apjl}{Astrophys. J. Lett.}%
\newcommand{\aap}{Astron. Astrophys.}%
\newcommand{\mnras}{Mon. Not. Roy. Astron. Soc.}%
\newcommand{\fpeak}{f_\mathrm{peak}}
\newcommand{\snr}{S/N}
\newcommand{\fspiral}{f_\mathrm{spiral}}
\newcommand{\mtot}{M_\mathrm{tot}}
\newcommand{\mthres}{M_\mathrm{thres}}

\newcommand{\heff}{h_\mathrm{char}(f)}

\newcommand{\ntrain}{N_\mathrm{train}}
\newcommand{\nval}{N_\mathrm{val}}
\newcommand{\ntest}{N_\mathrm{test}}
\newcommand{\ntot}{N_\mathrm{tot}}
\newcommand{\F}{\mathcal{F}}
\newcommand{\Fs}{\mathcal{F}{s}}
\newcommand{\Ov}{\mathcal{O}}

\newcommand{\Dl}{D_\mathrm{L}}

\newcommand{\apr}[1]{\operatorname{KNN}_{{\mathrm{tr}=#1}}^{{\mathrm{APR4}}}}
\newcommand{\knneos}[1]{\operatorname{KNN}_{{\mathrm{tr}=#1}}^{{\mathrm{EOS}}}}

\newcommand{\sfhx}[1]{\operatorname{KNN}_{{\mathrm{tr}=#1}}^{{\mathrm{SFHX}}}}

\begin{document}

\preprint{APS/123-QED}

\title{Gravitational-wave model for neutron star merger remnants with supervised learning}

\author{Theodoros Soultanis\,\orcidlink{0000-0002-7405-6184}}\email{t.soultanis@gsi.de}
\affiliation{GSI  Helmholtzzentrum  f\"ur  Schwerionenforschung,  Planckstra{\ss}e  1,  64291  Darmstadt,  Germany}
\author{Kiril Maltsev\,\orcidlink{0000-0001-8060-7416}}
\affiliation{Heidelberger Institut f{\"u}r Theoretische Studien,
        Schloss-Wolfsbrunnenweg 35, D-69118 Heidelberg, Germany}
\affiliation{Zentrum f\"ur Astronomie der Universit\"at Heidelberg, Institut f\"ur Theoretische Astrophysik, Philosophenweg 12, D-69120 Heidelberg, Germany}
\author{Andreas Bauswein\,\orcidlink{0000-0001-6798-3572}}
\affiliation{GSI  Helmholtzzentrum  f\"ur  Schwerionenforschung,  Planckstra{\ss}e  1,  64291  Darmstadt,  Germany}
\affiliation{Helmholtz Research Academy Hesse for FAIR (HFHF), GSI Helmholtz Center for Heavy Ion Research, Campus Darmstadt,  Germany}
\author{Katerina Chatziioannou\,\orcidlink{0000-0002-5833-413X}}
\email{kchatziioannou@caltech.edu}
\affiliation{Department of Physics, California Institute of Technology, Pasadena, California 91125, USA}
\affiliation{LIGO Laboratory, California Institute of Technology, Pasadena, California 91125, USA}

\author{Friedrich K.\ R{\"o}pke\,\orcidlink{0000-0002-4460-0097}}
\affiliation{Heidelberger Institut f{\"u}r Theoretische Studien,
        Schloss-Wolfsbrunnenweg 35, D-69118 Heidelberg, Germany}
\affiliation{Zentrum f\"ur Astronomie der Universit\"at Heidelberg, Institut f\"ur Theoretische Astrophysik, Philosophenweg 12, D-69120 Heidelberg, Germany}
\author{Nikolaos Stergioulas\,\orcidlink{0000-0002-5490-5302}}
\affiliation{Department of Physics, Aristotle University of Thessaloniki, 54124 Thessaloniki, Greece}

\date{\today}

\begin{abstract}
We present a {time-domain model} for the gravitational waves emitted by {equal-mass} binary neutron star merger remnants {for a fixed equation of state.}
We construct {a large} set of {numerical relativity} simulations for {a} single equation of state {consistent} with current constraints, totaling 157 equal-mass binary neutron star merger configurations. 
{The {gravitational-wave} model is {constructed using} the supervised learning method of K-nearest neighbor regression.} 
As a first step toward developing a general model with supervised learning methods {that accounts for the dependencies on equation of state and the binary mass{es} of the system,} we {explore the impact of the size of the dataset on the {model}.} {We assess the accuracy of the {model} for a varied dataset size and number density in total binary mass.} Specifically, we consider five training sets of $\{ 20,40, 60, 80, 100\}$ {simulations} {uniformly distributed in total binary mass}. We evaluate the {resulting} models in terms of {faithfulness} {using a test set of 30 {additional} simulations that are not used during {training} {and which are equidistantly spaced in total binary mass}}. {The models} {achieve {faithfulness} with maximum values in the range of {$0.980$ to $0.995$}}. We assess our {models} {simulating signals observed by} the three-detector network of Advanced LIGO-Virgo{. We find} that all {models} with training sets of size equal to or larger than $40$ {achieve an unbiased measurement of the main gravitational-wave frequency}. {We confirm that our results do not depend qualitatively on the choice of the (fixed) equation of state.} We conclude that training sets, with a minimum size of $40$ simulations, or a number density of approximately $11$ simulations per $0.1\,M_\odot$ of total binary mass, suffice for the construction of faithful templates {for the} post-merger {signal}{ for a single equation of state and equal-mass binaries}{, and lead to mean faithfulness values of $\F\simeq 0.95$.}  {Our model being based on only one fixed equation of state represents only a first step towards a method that is fully applicable for gravitational-wave parameter estimation. However, our findings are encouraging since we show that our supervised learning model built on a set of simulations for a fixed equation of state successfully recovers the main gravitational-wave features of a simulated signal obtained using another equation of state. This may indicate that the extension of this model to an arbitrary equation of state may actually be achieved with a manageable set of simulations.}
\end{abstract}

\maketitle
\section[Introduction]{Introduction}\label{introduction}

Two detections of gravitational-waves (GWs) from sources identified as binary neutron star {(BNS)} mergers, GW170817~\cite{2017PhRvL.119p1101A} and GW190425~\cite{2020ApJ...892L...3A}, have been {reported} already, and many more observations are anticipated in the next years~\cite{Aasi:2013wya}. The GWs emitted by BNS mergers provide a new channel for probing the properties of high-density matter in the interior of neutron stars. Already GW170817 {has} led to stringent constraints on the so far incompletely known neutron star equation of state (EOS)~\cite{2017PhRvL.119p1101A,LIGOScientific:2018hze,Abbott:2018exr}, by the measurement {of the tidal deformation of the binary components} of the inspiral phase (see~\citep{2020GReGr..52..109C,2021GReGr..53...27D} for reviews). Further constraints {on the EOS} could be derived from the electromagnetic counterparts of GW170817 or by combined measurements, such as those discussed in~\cite{Margalit_2017,Bauswein_etal_2017,PhysRevD.96.123012,Radice_2018,Ruiz2018,Rezzolla_2018,2020NatAs...4..625C,2020NatAs...4..625C,2020Sci...370.1450D,2020PhRvD.101l3007L,2020Sci...370.1450D,AB2021june,Raaijmakers2021sep,2021PhRvD.104f3003L,2021arXiv210508688P,2021MNRAS.505.3016N,Huth2022,Brandes2023,2024A&A...689A..51B} and the references therein. {In addition,} the increased number of detections expected in the near future would improve those {constraints}~\cite{DelPozzo2013,2015PhRvD..92j4008C,2015PhRvD..91d3002L,2019PhRvD.100j3009H,2020PhRvD.101d4019C,PhysRevD.107.043021}. The post-merger GW {signal} in GW170817 could not be detected because the sensitivity of the Advanced LIGO~\cite{LIGOScientific:2014pky} and Advanced Virgo~\cite{VIRGO:2014yos} detectors was not sufficient~\cite{2017PhRvL.119p1101A,2017ApJ...851L..16A,LIGOScientific:2018hze}. However, with upgraded second-generation~\cite{KAGRA:2020npa}, future next-generation~\cite{LIGOScientific:2016wof,Maggiore:2019uih,2019BAAS...51g..35R,Srivastava_2022}, or dedicated high-frequency detectors~\cite{2019PhRvD..99j2004M,Ackley:2020atn,2021PhRvD.103b2002G,2021CmPhy...4...27P,2021arXiv211010892S}, such detections are likely to be achieved in the next years. 

{In the post-merger phase, if the total binary mass is lower than a threshold mass ($\mthres$)~\cite{Hotokezaka2011,PhysRevLett.111.131101,AB2021june,Tootle_2021} for prompt black hole formation, the remnant is a quasi-stable rapidly rotating neutron star that is supported against gravitational collapse {by} differential rotation and thermal pressure. This object is typically referred {to} as a \textit{hypermassive neutron star} (HMNS) {in the literature}. The remnant oscillates in various fluid (quasi) oscillation modes and emits GWs in the range of 2-4~kHz. The most prominent  feature of the post-merger GW emission is associated with the fundamental quadrupolar oscillation mode (see~\cite{Zhuge1996,Shibata2005a,Shibata2005b,Oechslin2007,Stergioulas2011,Bauswein2012Jan,Bauswein2012Sep,Hotokezaka2013,Takami2015,Bernuzzi2015,Clark2016,AB2016,Foucart2016,Dietrich_2017,PhysRevD.96.063011,Liebling2021,Dudi2022}). The frequency of this mode, denoted as $\fpeak$ or $f_2$, depends on the EOS.} As a result, a measurement of the GWs emitted in the post-merger phase would provide substantial and independent constraints {to the EOS, see~\citep{Bauswein2019sep,Baiotti2019,Friedman2020,Bernuzzi2020,Diedrich2021mar,Sarin2021jun} and references therein}.

The detection of GWs emitted from BNS mergers relies on robust and efficient data analysis techniques. One approach is to employ matched filtering search methods that rely on post-merger GW models. To achieve this, reliable models in the time or frequency domain for the post-merger GW emission which are informed by numerical relativity simulations, also called templates, are required. {Several parametric models have been introduced already, either in {the} frequency domain~\cite{Messenger2014,PhysRevD.100.044047,2023PhRvD.107l4009P,2022arXiv220509112B} or in {the} time domain~\cite{Hotokezaka2013,AB2016,Bose2018,Yang2018,PhysRevD.100.104029,Easter2020,TS2022,Królak_2023}}. An alternative method employs morphology-independent {models}~\cite{PhysRevD.96.124035,PhysRevD.99.044014,PhysRevD.105.104019,2023CQGra..40v5008T,PhysRevD.107.043021,2023PhRvD.107j3053M,2023arXiv231110626S}.

In this work, {we focus on} {the construction of models informed by numerical relativity simulations}. The authors of Ref.~\cite{2022arXiv220106461W} introduced a time domain model employing a conditional variational auto-encoder (CVAE) using simulations {of BNS mergers in numerical relativity}. {The CVAE is a deep-learning method where two sets of variables are employed for the interpolation. The first set of variables characterizes the system. These variables can be the total mass, binary mass ratio, initial spins, etc. The second set of variables, that is, the latent variables, are not known a priori but the CVAE algorithm learns to use them during training. Latent variables include the EOS, discretization errors,  physics-modeling choices, etc.} The authors generated an extended data set using a stochastic model and estimated that training such CVAE models would require approximately $10^4$ signals. Reference~\cite{Clark2016} introduced a frequency domain model for the amplitude and phase of the spectra, based on principal component analysis for a set of numerical relativity simulations. The authors of Ref.~\cite{Easter2019} introduced a hierarchical frequency domain model, for the amplitude of spectra, that trains on existing numerical-relativity post-merger simulations. {They employed GW spectra aligned in frequency, as in~\cite{Clark2016}, and fitted the amplitudes with a linear model.} {In \cite{2024Pesios}, the amplitude of the post-merger spectrum was predicted using two different types of regression models in the frequency domain, multi-linear regression (MLR) and artificial neural network (ANN) regression, which were trained on a total of 87 simulations spread among 14 different EOS. With both regression methods, high fitting factors were achieved, when the method was calibrated to reduce the uncertainties inherent in empirical relations for the main post-merger frequency.}

{One of the main challenges in the construction of post-merger models informed by numerical relativity simulations is that they can be computationally costly, and thus it is difficult to create a sufficiently large library of simulations.} Furthermore, general-purpose GW models must take into account the various degrees of freedom that influence the properties of the GW signals, such as the total binary mass $\mtot=m_1+m_2$, the binary mass ratio $q=m_1/m_2$\footnote{$m_1,m_2$ correspond to the individual gravitational masses, at infinite separation, of the two companion neutron stars.}, the EOS model, intrinsic spin, etc. which significantly increase the size of the required data set of simulations. In this work, we vary the total binary mass $\mtot$ and fix the EOS model and the mass ratio to $q=1$. We intend to study the {other degrees} of freedom in future work. The current study allows to understand the minimum requirement for using supervised learning techniques and thus to assess the performance of such methods.

We introduce a {model} for post-merger GWs employing the supervised learning method of K-nearest neighbor (KNN) regression. {In this method, the assigned value at the interpolation point is computed based on the mean of the values of its K-nearest neighbors. Weights can be assigned to each point in the local neighborhood using arbitrary functions of distance in the parameter space.} We consider a sequence of equal-mass binaries with increasing total binary mass $\mtot$ for a fixed EOS model. We build {a large library of simulations} for a fixed EOS model, consisting of a total of 157 binaries {uniformly distributed with respect to the total binary mass of the system} using a smooth particle hydrodynamics  (SPH) code~\cite{SPH1,SPH2,PhysRevD.81.024012} (see Sec.~\ref{methods}).  We split the data set into three subsets used for training, validation, and testing, and optimize models with training sets of varied sizes. We assess {the KNN models, i.e., models constructed with KNN regression,} in terms of {a noise weighted inner product between two signals, {the faithfulness ($\F$)~\footnote{A perfect match between two signals corresponds to the value of $\F=1$.}}. }{In addition, we carry out injections, that mimic real detection scenarios, and enable us to evaluate our models. In those, a simulated signal with specific binary source parameters is inserted into a detector network and then reconstructed by the model using parameter estimation methods. We inject the simulated signals from the test set configurations and consider the Advanced LIGO-Virgo network.} We find that KNN models with training sets of size equal or larger than $40$ exhibit convergence, and {accurately} reconstruct the {dominant $\fpeak$ frequency of the} {injected} signals. The models achieve high values of {$\F$} with {mean values of approximately $0.95$ and} maximum values in the range of $0.992$ to $0.995$, for simulated signals of the test set.

Furthermore, we study the execution time of templates generated with the KNN method. {We find that the KNN models achieve execution times of approximately a few milliseconds for sampling rates comparable to those employed by Advanced LIGO.}

Finally, we {explore} the generality of our findings by constructing {another} KNN model {for a fixed EOS} using an alternative EOS model. We show that a training set with $40$ simulations suffices to build a reliable {time-domain} model that achieves high {mean} values of {$\F$} {in that case too. {In~\cite{2024Pesios} the same order of magnitude of number of simulations was used to create MLR and ANN-based regression models to predict the amplitude in the frequency domain.}}

This paper is structured as follows: In Sec.~\ref{methods} we describe the methods employed for the different stages of this work, i.e., the simulation tool and the parameters of the simulated systems, the employed supervised learning method, and the numerical setup of the injection studies. In Sec.~\ref{results} we present our results. We evaluate the performance of the various KNN models in terms of the faithfulness $\F$ achieved between the model predictions and the simulations of the test and validation sets. We present information regarding the speed of such type of models{ and discuss the results of the injection studies}. Finally, we construct a KNN model using a different EOS model and assess the effect of the EOS on the accuracy of the template and the necessary sampling density to achieve high performance. {We discuss our conclusions in Sec.~\ref{Conclusions}}.

\section[Methods]{Methods}\label{methods}

{In this section, we discuss the methods employed in this work.} {We} describe the numerical setup of the BNS simulations. We describe the main concepts of the KNN scheme, the preparation of data, and how the optimization of the KNN model and the hyper-parameter tuning is carried out. We discuss {the scheme for} splitting the original GW data set into subsets that are used for training, validation, and testing of the KNN model. We present the complete form of the model. {Finally, we describe the setup of our injection studies. }

\subsection{BNS merger simulations}

We perform numerical relativity BNS merger simulations using a general relativistic SPH code~\cite{SPH1,SPH2,PhysRevD.81.024012}. The spacetime is evolved under the conformal flatness condition (CFC)~\cite{CFC1,CFC2}. In this version of the code we use the Wendland kernel function~\cite{schaback_kernel_2006,rosswog_sph_2015}.

We simulate a sequence of {157} equal-mass binaries with increasing binary mass. We employ the APR4~\cite{Akmal:1998cf} EOS model. {For this EOS model, the thermal effects are approximated by an ideal gas pressure component with $\Gamma_\mathrm{th}=1.75$ (see e.g.~\cite{AB2010}).} The total binary masses of the sequence of simulations range from $2.4~M_\odot$ to $2.76~M_\odot$. {The simulations are uniformly distributed with respect to total binary mass $\mtot$ (see Fig.~\ref{fig:test_sets_complete_set.pdf})}. Because the threshold mass $\mthres$ for prompt black hole formation is $2.825~M_\odot$~\cite{AB2021june} for this EOS model, the chosen $\mtot$ range results in remnants that do not undergo a delayed collapse for at least a few tens of milliseconds. {From every simulation, we obtain the $+$ and $\times$ polarizations of the GW signal, denoted by $h_+(t; \mtot)$ and $h_\times(t; \mtot)$, for an observer at the rotational axis of the binary.} {The GW signal is artificially amplified by approximately $40\%$ to account for the quadrupole formula's underestimation of the GW luminosity (see~\cite{2005PhRvD..71h4021S}).} We then interpolate the signal with a sampling rate of $dt\approx 0.005~$ms (or $200~$kHz).


Furthermore, we construct a second mass-sequence of equal mass binaries using the SFHX EOS model~\cite{SFHX_SFHO}. The SFHX EOS model provides a consistent treatment of thermal effects. We simulate a total of {99} binaries with $\mtot$ ranging from {$2.4~M_\odot$ to $2.8~M_\odot$}. With this data set, {we construct another KNN model for a fixed EOS model, i.e. the SFHX EOS, and {discuss the agreement between the results obtained using} the two different EOS models}.

{To extract the $\fpeak$ frequency, for every simulation, we construct the complex signal defined as $h(t)=h_+(t) -i h_\times(t)$. We then compute the Fourier Transform of the post-merger GW signal using the implementation of Numpy~\cite{numpy}. We use the Tukey window function with a roll-off parameter of 0.05. Then we obtain the characteristic frequency-domain GW strain defined as $\heff=\tilde{h}(f)\cdot f$, where $\tilde{h}(f)$ is the Fourier Transform of $h(t)$. We identify $\fpeak$, i.e., the frequency of the quadrupolar oscillation mode, as the frequency corresponding to the most dominant peak in the GW spectrum. Finally, a second-order polynomial fit is performed in the local region of this peak to determine the exact $\fpeak$ value at which the maximum occurs.}

\subsection{K-nearest neighbours model}\label{K-nearest neighbours template}

K-nearest neighbor regression is a supervised learning method {that allows for non-parametric fitting}. {In this work, we use KNN regression to describe the emission of GWs in the post-merger phase using training sets of different sizes (number of simulations).} The goal is to predict GW signals of unseen configurations. {As is typically done in other supervised learning methods, we train the model using the strain data from the training set.} {{In addition, we aim} to estimate the size of the dataset that is required for building accurate post-merger models. {In this regard, the KNN regression is a simple yet robust method that allows us to explore concepts of supervised learning regression without dealing with the complexity of more sophisticated schemes such as deep learning approaches. Furthermore,} a relatively smooth transition between the GW spectra with respect to $\mtot$ is expected as in our model the EOS is fixed. Thus, regression schemes that use information from neighbors may be appropriate to use.} For our analysis we use the KNN implementation provided by SciKit~\cite{scikit-learn}.

\subsubsection{Definitions}

{The overlap ($\Ov$) is a widely used metric that quantifies the match between two signals. Perfect agreement corresponds to an $\Ov$ value of $1$. The overlap $\Ov$ is defined as} 
{\begin{eqnarray}\label{Fitting factor definition}
\Ov &\equiv& \frac{\bigl< h_1(t), h_2(t)\bigl>}{\sqrt{\bigl< h_1(t), h_1(t) \bigl>  \bigl< h_2(t), h_2(t) \bigl>}}\,.
\end{eqnarray}

The noise-weighted inner product $\left\langle h_1(t), h_2(t) \right\rangle$ between the signals $h_1(t)$ and $h_2(t)$ is given by
\begin{eqnarray}\label{noise-weighted product definition}
\bigl< h_1(t), h_2(t)\bigl> &\equiv& 4 \mathrm{Re} \int_0^\infty df \frac{\tilde{h}_1(f)\cdot \tilde{h}_2^*(f)}{S_h(f)}~,
\end{eqnarray}
{where $S_h(f)$ is the {one-sided} noise spectral density of the detector {of Advanced LIGO at design sensitivity~\cite{LIGOScientific:2014pky}}, and $\tilde{h}_\mathrm{1}(f)$, $\tilde{h}_\mathrm{2}(f)$ the Fourier transforms of $h_1(t),~h_2(t)$.}


The maximized overlap $\Ov$ with respect to initial phase $\phi_0$ and merger time $t_0$ defines the {faithfulness} $\F$ (or {match}) between two signals. It reads
\begin{eqnarray}\label{Match}
\F &\equiv&  \max_{\phi_0, t_0}~\frac{\bigl< h_1(t), h_2(t)\bigl>}{\sqrt{\bigl< h_1(t), h_1(t) \bigl>  \bigl< h_2(t), h_2(t) \bigl>}}\,.
\end{eqnarray}
{In what follows, $\Ov$ and $\F$ are calculated only for the post-merger signals ($t>t_\mathrm{0}$) while the inspiral is not used (see below). We consider a frequency band, for Eq~\eqref{noise-weighted product definition}, from $10$~Hz to $5.0$~kHz.}

{We assess how well the model reproduces the simulated signals in terms of faithfulness $\F$, i.e., the maximized in phase and time overlap, between the simulations and the model's predictions for the same value of $\mtot$ (see Sec.~\ref{Performance of KNN models} and Sec.~\ref{EOS dependence}). For the injections (see Sec.~\ref{Injections}), that correspond to the realistic parameter estimation scenario, all parameters of the model are varied (including $\mtot$). In this regard, the overlap $\Ov$ distribution shows how well the reconstructed signal matches the injected signal. }

\subsubsection{K-nearest neighbors regression}
{
We summarize the general concepts of the KNN regression (see~\cite{james2023introduction}). We consider a point $\vec{X}_0$ in the N-dimensional {regressor} parameter space of the interpolation problem and the corresponding {target variable} prediction denoted by $\vec{Y}_0$. The algorithm first identifies the $K$ points of the training set that are closest to $\vec{X}_0$, denoted by $\mathcal{N}_0$. The prediction $\vec{Y}_0$ is then estimated using the average (or weighted average) of all points in $\mathcal{N}_0$. The prediction estimate is given by the general form 
\begin{eqnarray}
        \vec{Y}_0 &=&\frac{1}{K} \sum\limits_{\vec{X}_i \in \mathcal{N}_0}  \dfrac{w_i\cdot\vec{Y}_i}{W},\quad\mbox{where}~W=\sum\limits_{\vec{X}_i \in \mathcal{N}_0} w_i\,,
\end{eqnarray}
where $\vec{Y}_i$ are the observations i.e., the strain data in our case, of the training set, and $w_i$ are assigned weights between $\vec{X_0}$ and the points in $\mathcal{N}_0$. 

The weights can be chosen to be equal for all neighbors or according to arbitrary criteria. For example, an inverse proportionality of the form $w_i=1/d_i$ where $d_i$ is the distance between $\vec{X}_0$ and the neighbors $\vec{X}_i$ in $\mathcal{N}_0$ is commonly used. Furthermore, the metric that defines the calculation of distances between points in the N-dimensional space {need to be} chosen {before the training}. }

{The number of neighbors $K$, the weighting criterion, or {the metric which determines} how distances in the parameter space are measured, are not known a priori and are considered tuning parameters, commonly referred as \textit{hyper-parameters} (see SciKit~\cite{scikit-learn} and references therein). The hyper-parameters affect the accuracy and performance of the KNN regression model. For this reason, we use the validation data to determine the optimal set of hyper-parameters. This stage is called \textit{hyper-parameter tuning}.}

{In order to determine the optimal set of hyper-parameters, for each training set, we evaluate the accuracy of the KNN regression predictions for the simulations of the validation sets by considering a global quantity that is representative of the
overall performance.} We opt for the mean value of faithfulness {$\F$}, defined in Eq.~\eqref{Match}, between the signals of the validation set and the corresponding predictions of the various KNN models. {{A possible caveat of this choice of global quantity may be that the late part of the signal, where the amplitude is low, is penalized less than the early part.}} {For each training set, we iterate over all combinations of hyper-parameters pre-defined over a grid, and store the set of hyper-parameters that maximize the mean {$\F$} over the validation data. 

In our case, the input features{, i.e., the space where the vectors $\vec{X}_i$ live,} of the KNN model are the total binary mass $\mtot$ and the {simulation} coordinate time $t$ {expressed as an array with sampling rate $f_\mathrm{s}=1/dt$}. {The vectors $\vec{Y}_i$ are $+$ polarization simulated signals $h_+(t; \mtot)$  expressed as arrays of the same length as $t$}. The output of the KNN model is a GW signal {for the $+$ polarization}. 

\subsubsection{The model}

{We find that the KNN algorithm performs best when the input GW signals used in training are aligned in phase.} {In order to align the signals in phase, we express the GW signal in the complex form
\begin{eqnarray}
h(t)&=&h_+(t) + i~h_\times(t)\\
    &=&|h(t)|\cdot e^{+i \phi(t)}\,,  
\end{eqnarray}
where the amplitude $|h(t)|$, and phase $\phi(t)$ are defined as
\begin{eqnarray}
    |h(t)|&=&\sqrt{h_+^2(t)+h_\times^2(t)}\,, \\
    \phi(t)&=&  \tan^{-1}\left(\frac{h_\times(t)}{h_+(t)}\right)\,.
\end{eqnarray}
The merger time $t_\mathrm{merge}$ is the time at which $|h(t)|$ reaches its maximum value, and the phase during the merger time is given by $\phi_\mathrm{merge}=\phi(t_\mathrm{merge})$. We then introduce a phase shift $\phi\rightarrow \phi-\phi_\mathrm{merge}$, and reconstruct the two polarizations as
\begin{eqnarray}
    h_+(t) &=& \mathrm{Re}(h(t))\,, \\ 
    h_\times(t) &=& \mathrm{Im}(h(t))\,. 
\end{eqnarray}
\begin{figure}[h!]
	\includegraphics[scale=0.33]{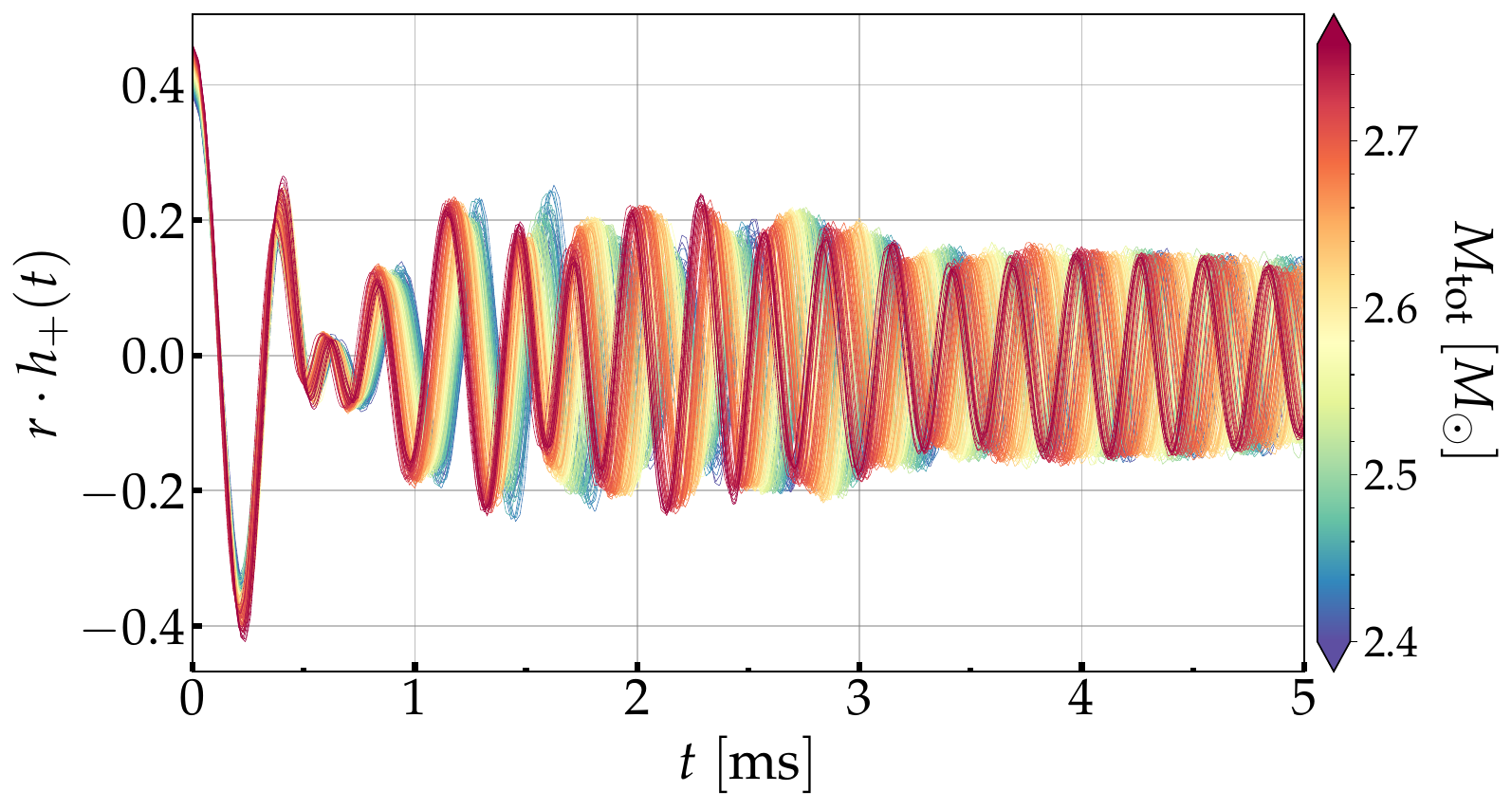}
	\caption{\label{fig:GWa_aligned.pdf}Aligned GW signals for the complete set of 157 simulations. The colors indicate the different total binary mass. The first $5~$ms of the GW signals are shown, while the total duration of the signals is 17~ms. }
\end{figure}
We also introduce a time shift $t\rightarrow~t-t_\mathrm{merge}$ so that $t=0$ corresponds to the merger time}. Figure!\ref{fig:GWa_aligned.pdf} shows the aligned signals $h_+(t)$ for the complete set of simulations where the color code indicates the total binary mass $\mtot$. Interestingly, we find remarkable agreement in the morphology of the aligned signals in the {first 1~ms}. In the early phase, {up to  1~ms}, the signal is simple, as it exhibits almost no dependence on the total binary mass $\mtot$. The dependence on $\mtot$ becomes apparent as the different signals start to fall out of phase. This is because the characteristic frequencies of the GW signal scale {proportional} to the total binary mass of the system. Phase alignment simplifies the first {1~ms} of the post-merger signal and ensures that it is better {modeled} by the KNN model than in the case without phase alignment.

Finally, we discuss the general form of our model in the binary frame, which is numerically equivalent to a face-on observer. {The antenna response functions of the detectors for a source with arbitrary position and orientation in the sky can be computed using the detector projection tensors~\cite{MAGGIORE2000283,MaggioreVol1} which introduce several parameters: a) the right ascension ($\mathrm{r.a.}$) and the declination ($\delta$), which describe the location of the source in the sky; b) the  inclination angle {of the binary's orbital plane} ($\iota$); {and} c) the polarization angle ($\psi$), which is an angle in the plane perpendicular to the line of sight that defines the $+$ and $\times$ polarizations.} {In order for the KNN model's predictions to describe general GW signals, we also introduce an appropriate phase shift of the form $e^{i \phi_0}$.} {This} phase shift is typically {used} in every detection search scenario, since GW signals have arbitrary initial phases depending on the position of the detector and arrival of the signal (see Sec.~\ref{Injections}). {In order to account for the arbitrary arrival time of the GW signal we also introduce a time shift parameter $t_0$ which corresponds to the start of the post-merger signal, or equivalently to the merger time.} The model reads
\begin{eqnarray}
h_+(t)      &=& \mathrm{KNN}(t-t_0; M_\mathrm{tot})~e^{i\phi_0}\label{KNN model}\,,\\ 
h_\times(t) &=& h_+(t-t_0)~e^{i\frac{\pi}{2}}\,,
\end{eqnarray}
{where KNN corresponds to the KNN model, and $t_0$ is the starting time of the post-merger signal.}

{Throughout the training and validation of the KNN model,} we impose a cut-off at $t=17~$ms of post-merger signal. {The KNN model produces longer post-merger signals than the cut-off time. The extrapolated signal depends on the data close to the cut-off region.} 

\subsection{Splitting of data}\label{Splitting of data}

{We split the APR4 data set (157 simulations) into subsets used for \textit{training, validation, and testing} of the models. The test set contains $\ntest=30$ simulations that we set aside and use for testing. The remaining data are split into training and validation sets. }

{The training set includes the simulations used to obtain KNN models for given sets of hyper-parameters. The validation set is used for the optimization of the KNN models, that is the hyper-parameter tuning, where the {set of hyper-parameters that maximizes the mean {$\F$}, is determined.} As we want to quantify the number of simulations required to build accurate models, we vary the size $\ntrain$ of the training set (the number of simulations for training) in the range of $\ntrain=\{20,40,60,80,100\}$. The size of the validation set is thus $\nval=N_\mathrm{tot} - \ntest - \ntrain$ where $\ntot$ is the total number of simulations. This split thus results in a total of five KNN models labeled by the size of their training set and the EOS, e.g. $\knneos{\mathrm{X}}$ where X denotes the size of the training set.}

{The hold-out test set provides an independent way to assess the performance of KNN models (with varying $\ntrain$) under the same conditions. The simulations of the test set are distributed in an approximate equidistant manner with respect to $\mtot$. }

{We employ the Latin Hypercube Sampler (LHS)~\cite{ef76b040-2f28-37ba-b0c4-02ed99573416} to draw $\ntrain$ simulations from a uniform distribution to construct the training set. LHS is a method for generating a homogeneously distributed sequence of parameter values using a randomized sampling scheme with memory. The advantage of LHS over the memoryless uniform random sampler is that it prevents formation of sampling gaps. These arise for the uniform random sampler as a statistical effect in consequence of grid point clustering, when the sample size is small~\cite{ef76b040-2f28-37ba-b0c4-02ed99573416}. The validation set contains the remaining simulations.}

\subsection{Injection setup}\label{Setup of the injection study}

We describe the setup for the injection study (see Sec.~\ref{Injections}). We perform injections using Bilby~\cite{Ashton_2019,2020MNRAS.499.3295R}, an inference code implemented in Python, which provides an infrastructure for GW parameter estimation. In this work, we use the Dynesty sampler~\cite{10.1093/mnras/staa278,sergey_koposov_2023_8408702,2004AIPC..735..395S,10.1214/06-BA127,2019S&C....29..891H}, implemented in Bilby\footnote{{The package version \texttt{2.1.1.} is used.}} (see~\cite{Ashton_2019} and the references therein for other sampler options), and consider 1000 live points. {For the auxiliary parameters of the Dynesty sampler, i.e., the number of autocorrelation lengths in the Markov-chain Monte Carlo algorithm and the number of walks taken by the live points, we choose values 100 and 3, respectively. We do not find a significant difference in the outcome of the injections using slightly different values}. We evaluate the performance of the KNN models by comparing the source parameters {of the reconstructed signals} {to those of the injected signals}. {We carry out several injections with varied luminosity distance ($\Dl$) of the source, and sample the parameters $\mtot,~\phi_0,~t_0,~\Dl$}.
{We use the posterior probabilities of the model's parameters and convert them to posterior probabilities of quantities connected to the the post-merger signal (see Sec.~\ref{Injections}). In this way, we examine the posterior probabilities of the dominant frequency peak, $\fpeak$, in the GW spectra and compare them to the values of the injected simulated signals. In addition, we discuss the posterior probabilities of {the overlap $\Ov$ values} between the reconstructed and the injected signals.}

We inject a GW signal for a binary configuration from the test set with a total mass of $2.656~M_\odot$ {in the source frame} and vary the luminosity distance $D_\mathrm{L}$ of the source from $1~$Mpc to $40~$Mpc. We use the three-detector Advanced LIGO Hanford, Livingston, and Advanced Virgo network, at design sensitivity. For the sky location of the source, we choose the combination of $\mathrm{r.a.}$ and $\delta$ which corresponds to  optimal orientation. {We consider a face-on binary, i.e., $\iota=0$.} {We do not include a noise realization in our analysis}. We employ uniform priors for the intrinsic parameters $\mtot$, $\phi_0$, $t_0$ in the range of $[2.4\,M_\odot, 2.8\,M_\odot]$, $[0.0, 2\pi]$, $[t_0^\mathrm{inject}-5\,\mathrm{ms}, t_0^\mathrm{inject}+5\,\mathrm{ms}]$, respectively\footnote{$t_0^\mathrm{inject}$ represents the injected $t_0$ value.}. For $\Dl$, we employ a logarithmic uniform prior in the range of $[0.48~\mathrm{Mpc}, 48~\mathrm{Mpc}]$. We use the sampling rate of Advanced LIGO, i.e., $f_s = 16384~\mathrm{Hz}$, and consider a signal of total duration of $1~$second {(see~\cite{PhysRevD.96.124035})}.

\section[Results]{Results}\label{results}
\begin{figure*}[t!]    
	\includegraphics[scale=0.26]{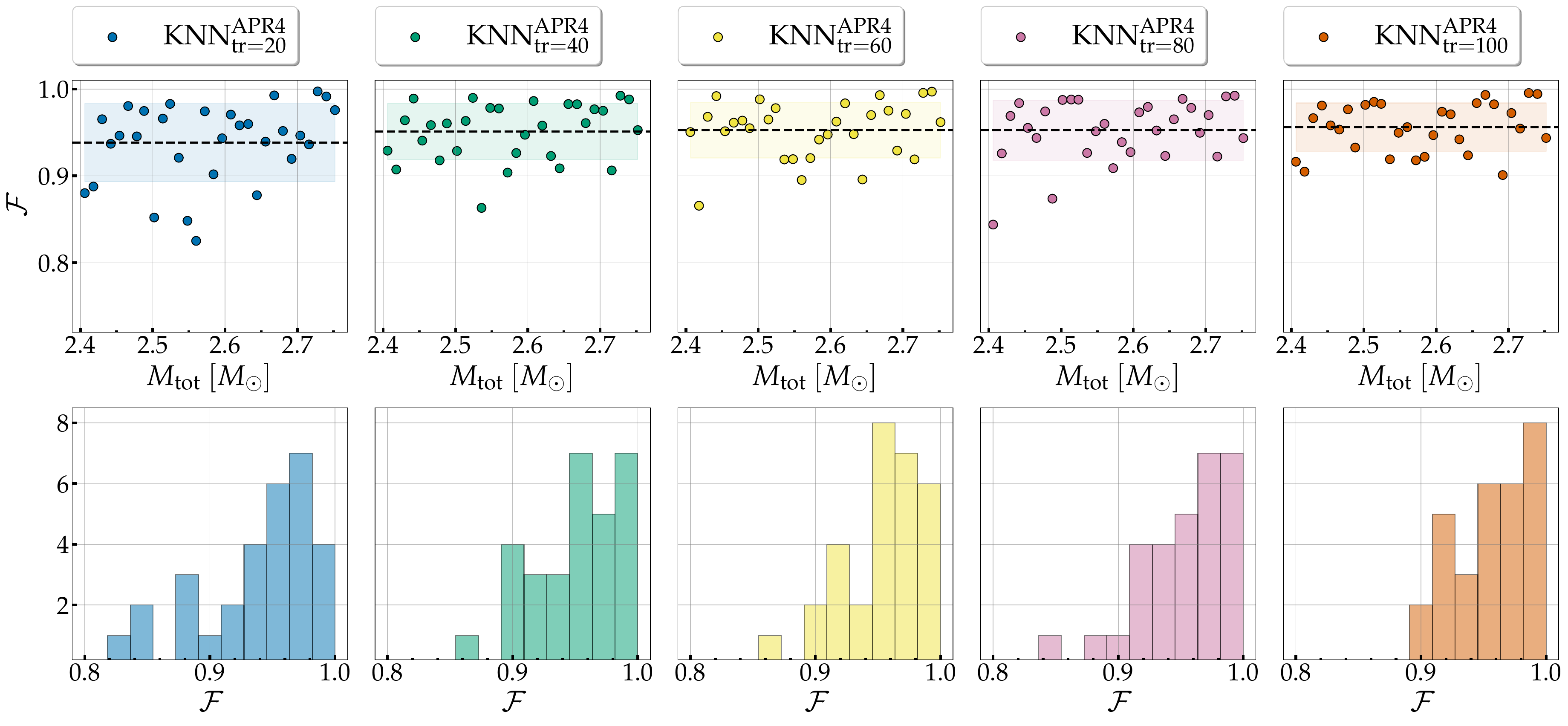}
	\caption{\label{fig:FFs.pdf} Upper panel: Faithfulness $\F$ as a function of the total binary mass $\mtot$ for the simulations of the test set and the predictions of the {different} KNN {models, where training sets of sizes $\ntrain=\{ 
 20,~40,~60,~80,~100\}$ were employed.} Black dashed line indicates the mean value. Shaded areas indicate the scatter of the data defined by the standard deviation. Bottom panel: Histograms of the aforementioned $\F$. }
\FloatBarrier
\end{figure*}

\subsection{Performance of KNN models}\label{Performance of KNN models}

{We assess the quality of the fits by calculating the values of {$\F$} between the simulated signals of the test set and the {signal} predictions {obtained from} the KNN models. We consider a known starting time of the signal of $t=0$.} {Because the simulated GW signals have arbitrary initial phases, we perform phase maximization with the} \texttt{SciPy} curve-fit routine~\cite{2020SciPy-NMeth}{, that employs a non-linear least squares algorithm}~\cite{TRF1,TRF2}. This procedure is repeated for {each of} the GW signals of the test set. The test set is a unique collection of configurations equidistantly distributed with respect to $\mtot$, and thus allows us to compare the different KNN models on the same GW signals. {In Appendix~\ref{Validation data}, we present the distribution of the {faithfulness {$\F$}}, computed using the above procedure, for the simulations of the validation set that were used during the hyper-parameter tuning of the KNN models.}

{We present} the {values of $\F$} between the predictions of the {various} KNN models, for each $\ntrain$, and the 30 simulations of the test set in Fig.~\ref{fig:FFs.pdf}. The upper row shows $\F$ as a function of the total binary mass $\mtot$, and the lower row shows the distribution of these $\F$ values. We find very good performance for most KNN models with almost no dependence on the size of the training set. {More specifically, a minimum of $\ntrain=40$ simulations used for training is sufficient to achieve high {$\F$}, and larger values of $\ntrain$ do not significantly enhance the overall performance of the model.}

As shown in Fig.~\ref{fig:FFs.pdf}, the mean value of $\F$ (indicated by a black dashed line) for most KNN models, i.e. $\apr{40}$ to $\apr{100}$, is approximately constant at $0.95$, and only marginally increases with $\ntrain$. These mean values are {comparable} to those of{~\cite{Easter2020}, where {$\F$} in the range of 0.92 to roughly 0.97 are reported for 9 simulated signals}. The mean value of the model $\apr{20}$, $\F=0.938$, is lower than that of the other models, however, it is still well above $0.90$. We also report the standard deviation, with respect to the mean value, for each KNN model (indicated by the shaded areas). The scatter is roughly constant for most KNN models except for the low-resolution $\apr{20}$ model where it is slightly larger than the rest. This may not be surprising since the overall performance of the KNN models with $\ntrain\geq 40$ converges (see Fig.~\ref{fig:FFs.pdf}). Furthermore, the best reproduced configurations achieve significantly larger {$\F$} values, close to perfect match ($\F=1$), {considering that the test set consists of signals that are not used in training or validation}. We do not find a clear trend in {$\F$} as a function of $\mtot $(see Fig.~\ref{fig:FFs.pdf}) for any of the KNN models.

The lower row of Fig.~\ref{fig:FFs.pdf} shows the distribution (histograms) of {$\F$} for the configurations of the test set. The KNN models {with large} $\ntrain$ exhibit a large number of counts at {the last two bins with the highest {$\F$} (in range of approximately 0.96 to 1.0)}. In particular, the $\apr{100}$ model corresponds to no counts below the histogram bin with center at $0.90$ {(with minimum {$\F$} at 0.901)}. {Quantitatively}, the difference between the models $\apr{40}$ to $\apr{80}$ is negligible with respect to the accuracy of the predictions. In practice, the model $\apr{40}$ is preferable, as it is computationally less expensive to construct a training set of simulations of size $\ntrain=40$ compared to $\ntrain=100$. 

In addition, we examine the outliers with the lowest {$\F$}. As shown in Fig.~\ref{fig:FFs.pdf}, for most KNN models ($\ntrain\geq 40$), the lowest values of {$\F$} are roughly around $0.85$. The mean value of {$\F$} (see above) represents the average error assigned to a prediction of the KNN models. 

Figure~\ref{fig:dumpGW_model} displays the GW signal obtained from the $\apr{100}$ model compared to the simulated signal from the test set for the configuration with $\mtot=2.728~M_\odot$. This configuration corresponds to the best performing prediction, with $\F=0.995$, of the test set. The agreement between these signals is remarkable, as indicated by the high {$\F$}. {This large {$\F$} is a consequence of our large data set as every configuration of the test set has close neighbors with respect to $\mtot$. In this particular case, the test simulation with $\mtot=2.728~M_\odot$, has closest neighbors in $\mtot$ from both sides, $\mtot=2.732~\mtot$ and $\mtot=2.726~\mtot$.} In Appendix~\ref{Best predictions for the unseen signals}, {we present the five best and five worst predictions for the simulations of the test set for the KNN models with $\ntrain=40,~100$ (see Fig.~\ref{fig:dumpGW_model_best_worst_040.pdf} and Fig.~\ref{fig:dumpGW_model_best_worst_100})}. Overall, we find very good agreement between the initial parts of the two signals (first 10~ms) which is important for achieving large values of {$\F$}. {Approximately in the first 5-7 ms, the signal} also contains information on the secondary frequency components of the spectrum{, apart from the dominant component,} and thus it is important for resolving those.

In the later parts of the signals, on average, all KNN models, regardless of $\ntrain$, tend to overestimate the decay of the signal (see also Fig.~\ref{fig:dumpGW_model_best_worst_100}). This effect may be a consequence of the chosen global metric, i.e., mean $\F$, used for the selection of the best-performing KNN models during the hyper-parameter tuning. As the late part of the signal is low in amplitude compared to the early part, leniency in fitting performance of the late part is not penalized as much as that of the early part. We also note that configurations neighbouring in $\mtot$ may exhibit small stochastic fluctuations in the decay times of their respective signals, which are triggered by numerical errors, and thus the KNN algorithm cannot resolve those trends over local neighborhoods. Although this systematic trend should be further investigated in the future, we find this effect to be negligible, since the {$\F$} of the configurations of the test set are already high.

{In summary, the various KNN models reproduce exceptionally well the GW signals of the test set with high precision ({$\F$} close to 1) in the whole $\mtot$ range. Our results suggest that KNN models with training sets of size $\ntrain \geq 40$ lead to accurate predictions for test set GW signals with large values of the {$\F$}, i.e., mean values roughly around $0.95$. The mean {of} {$\F$} appears to converge as the performance of the various KNN models is qualitatively the same for $\ntrain \geq 40$. However, it should be noted that the {$\F$} is a global metric for the match between two signals, and thus {depends on the total length of the signals.}

\subsection{Execution time of the model}

\begin{figure}[h]    \label{dumpGW_model}
 \includegraphics[scale=0.34]{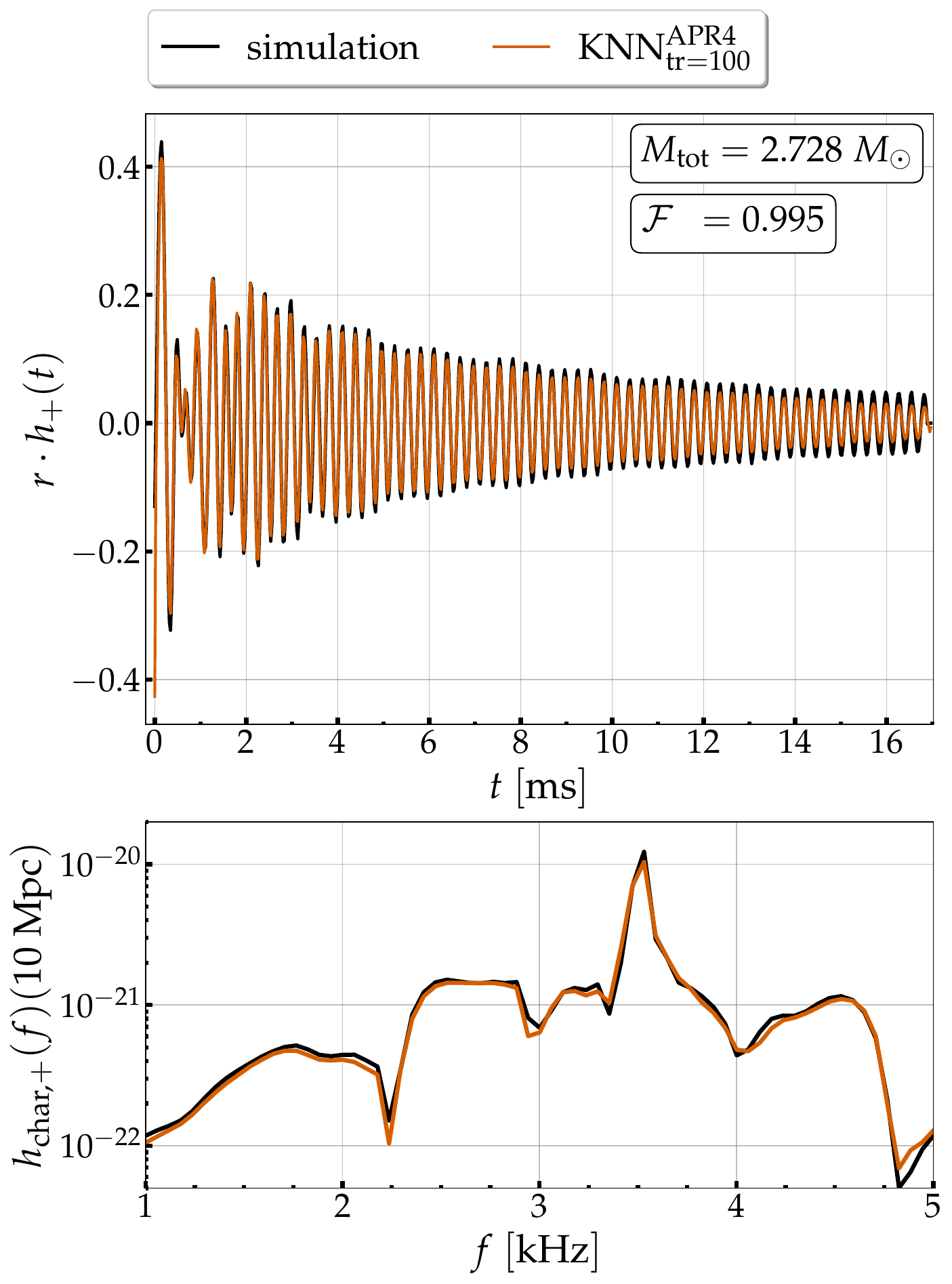}
 
	\caption{\label{fig:dumpGW_model} {Upper panel}: Comparison between the {time-domain} GW signal {for a simulation from the test set} (black) and the respective $\apr{100}$ model's prediction (orange).~Bottom panel: The same comparison but for the effective strain $\heff$. }
\end{figure}

Furthermore, we examine the {average} execution time of the KNN models, {i.e., the average time required to generate a signal with a duration of 17~ms}. As the goal is to use post-merger templates based on a KNN method (or other supervised learning methods) in real {parameter estimation} efforts, the execution time of the model must be reasonably short. This is because in inference, the model can potentially be called thousands of times. 

The speed of the model is largely dominated by the sampling rate used for the generated GW signal. As the sampling rate decreases (increases), the coordinate time $t$ array becomes less dense (more dense), and thus the KNN model generates fewer (more) data points. {This results in a shorter (longer) execution time for a call as fewer (more) data points need to be computed}. 
{To evaluate the speed of the model, we generate signals for every configuration of the test set and vary the sampling rate $f_s={1}/{dt}$. For each sampling rate we repeat the procedure a thousand times and store the mean value of the execution times.}

Figure~\ref{fig:speed_of_dt_of_all.pdf} displays the {average} 
execution  time for a single call for all KNN models. This corroborates the above statement that the execution time is an increasing function of the sampling rate. For reference, the black dashed line depicts the sampling rate for Advanced LIGO (aLIGO) {data}, $f_s = 16384~\mathrm{Hz}$. The execution time for all KNN models ranges from $1-3~\mathrm{ms}$ for sampling rates comparable to those of aLIGO. We note that the {execution} time reaches values of $\sim 10~\mathrm{ms}$ for unrealistically high sampling rates. This result is very encouraging, as it shows that templates constructed with supervised learning schemes can be employed {for parameter estimation, which requires million of waveform evaluations.}


\begin{figure}[h!]    
	\includegraphics[scale=0.34]{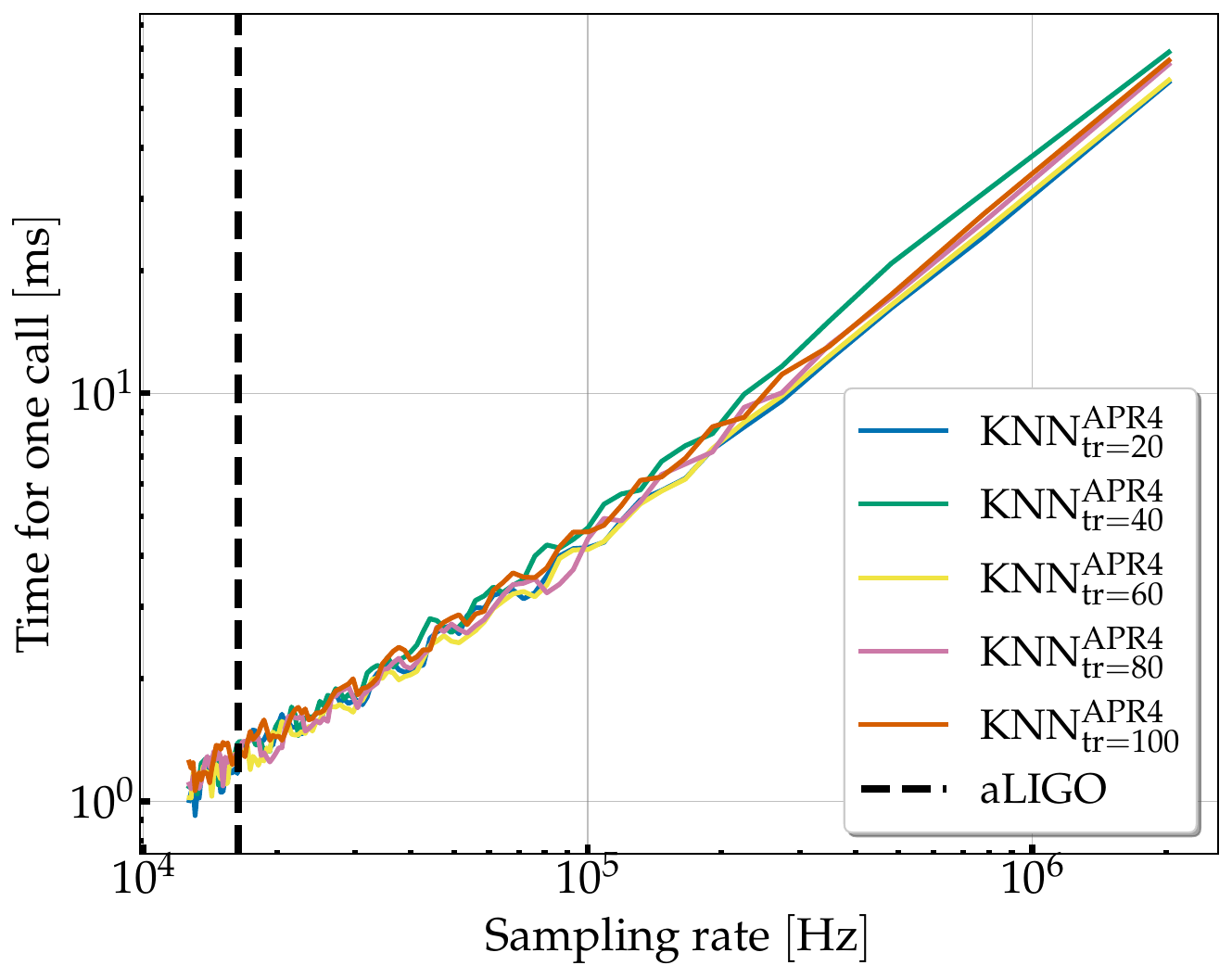}
\caption{\label{fig:speed_of_dt_of_all.pdf} Average execution time for one call of the KNN models as function of the sampling rate. Dashed black line indicates the sampling rate of Advanced LIGO. }
\end{figure}

Interestingly, as shown in Fig.~\ref{fig:speed_of_dt_of_all.pdf}, all KNN models exhibit approximately identical execution time curves, implying that the size of the training set does not influence the speed of the templates. This {may be} explained by the fact that for all KNN models, the \textit{number of neighbors} hyper-parameter (determined during hyper-parameter tuning) is approximately equal ($\sim 8-14$). {This is encouraging since a full EOS-dependent model will obviously rely on a large training set.}

\begin{figure*}[t!]    
	\includegraphics[scale=0.345]{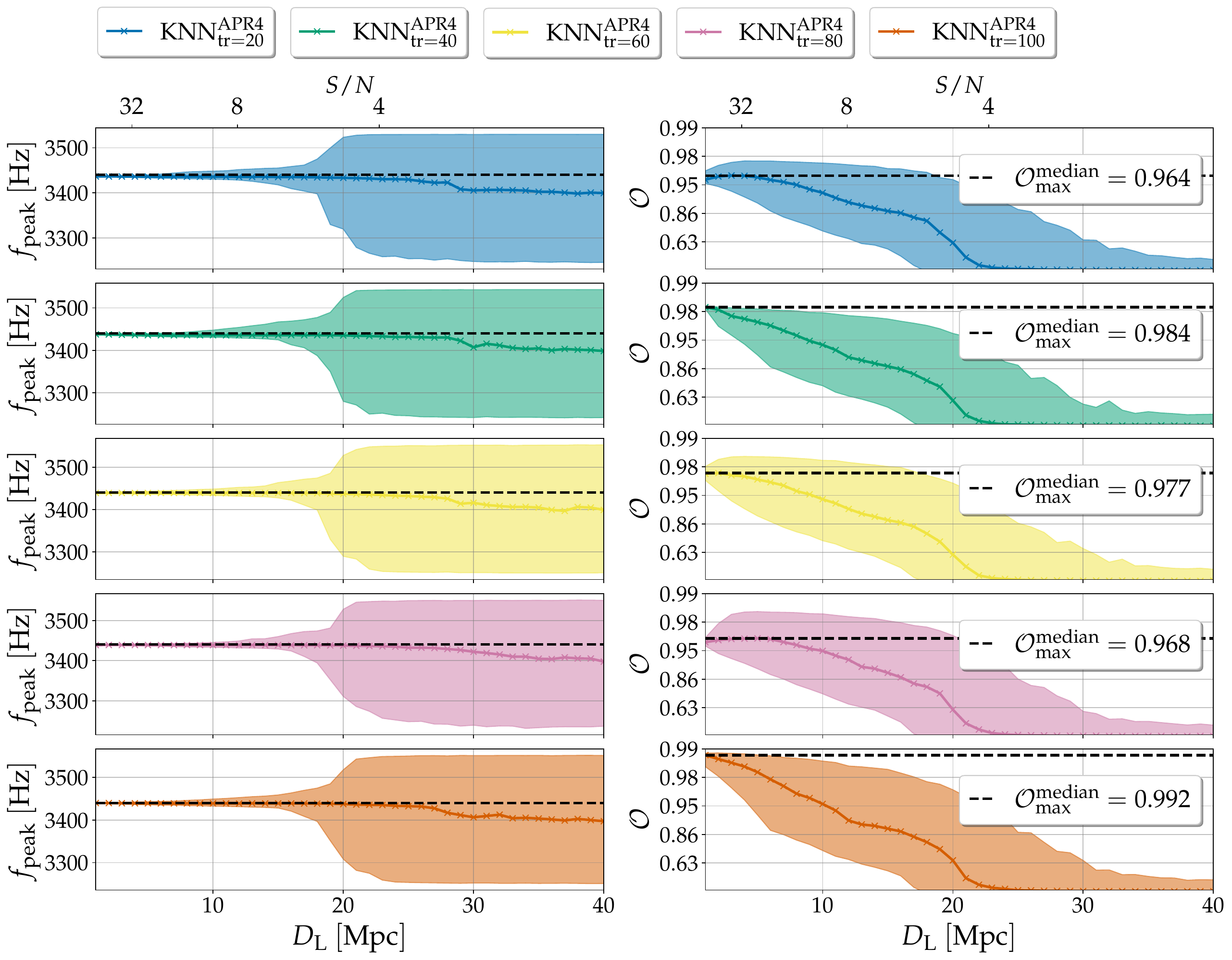}
\caption{\label{fig:injections_fpeak_recon_and_FF.pdf} Left panel: {Posterior} of $\fpeak$ for the simulation of the test set with $\mtot=2.656\,M_{\odot}$ using the KNN models as a function of the luminosity distance $\Dl$ {of the simulated signal}. Black dashed lines indicate the $\fpeak$ value of the injected signal. Colored curves show the medians, while shaded areas correspond to the $90\%$ credible intervals of the posterior probabilities. Right panel: As in the left panel, but for the corresponding $\Ov$ values. {Black dashed lines indicate the {maximum} median overlap $\Ov$ of the injected signal}. A three-detector network of Advanced LIGO and Advanced Virgo {at design sensitivity} is considered. The upper horizontal axis in both panels displays the corresponding $\snr$.}
\end{figure*}

\subsection{Parameter estimation on simulated data}\label{Injections} 

We discuss the results of our injection study for the configuration of the test set with the total binary mass $\mtot = 2.656~M_\odot$ {and varied luminosity distance} (see Sec.~\ref{Setup of the injection study}). {The three-detector network of Advanced LIGO and Advanced Virgo at design sensitivity is
employed.} {For each injection, {we sample the multi-dimensional posterior for} {$\mtot$, $\phi_0$, $t_0$, and $\Dl$}, from which the distributions of $\fpeak$ and {$\Ov$} are derived. To achieve this, we iterate through all {samples} {$(\mtot, \phi_0, t_0, \Dl)_i$} of the posterior distribution and generate GW signals $h_+(t; M_\mathrm{tot,i}, \phi_\mathrm{0,i}, t_\mathrm{0,i}, D_\mathrm{L,i})$ using the KNN models. We then take the Fourier Transform of the GW signals, extract $\fpeak${, as described in Sec.~\ref{methods}, } {for all samples}, and thus obtain the posterior distribution of $\fpeak$. 
Finally, we derive the posterior distribution of $\Ov$ by computing the match between {each reconstructed
GW sample} and the GW signal that corresponds to the injected parameters $(\mtot, \phi_0,t_0,\Dl)$.

Figure~\ref{fig:injections_fpeak_recon_and_FF.pdf} shows the  {90\% credible interval for} $\fpeak$ (left) and the corresponding $\Ov$ (right) as a function of the luminosity distance of the injected signal for the various KNN models. The black dashed lines correspond to the $\fpeak$ {and $\F$} of the injected simulated signal. As the luminosity distance of the injected signal decreases, the signal-to-noise ratio ($\snr$)~\cite{1994PhRvD..49.2658C} increases and thus the ability to reconstruct the signal is enhanced. This can be seen in Fig.~\ref{fig:injections_fpeak_recon_and_FF.pdf} where the uncertainties in $\Ov$, and also in $\fpeak$, become smaller at low $\Dl$ values. 

We find remarkable agreement between the KNN models with $\ntrain \geq 40$, as all of these models successfully reconstruct the injected signal and achieve high overlaps of approximately $\Ov\simeq 0.969-0.992$. {As $\Dl$ decreases, }the posterior distributions of $\fpeak$ and $\Ov$ become narrower, thus exhibiting reconstructed values of $\fpeak$ and $\Ov$ with smaller {uncertainty} at $90\%$ credible intervals. The $\fpeak$ medians, indicated by the {solid} curves in Fig.~\ref{fig:injections_fpeak_recon_and_FF.pdf}, approach the corresponding {true} value of the injected signal. As expected, we observe a drop of the {$\Ov$} as a function of $\Dl$. At a luminosity distance of roughly {${\Dl=18~}$Mpc}, the median $\Ov$ continues to be relatively high, i.e., $\Ov\simeq 0.80$, but drops below $0.60$ at approximately {${\Dl=22~}$Mpc}. The $90\%$ uncertainty in the reconstructed $\fpeak$ only becomes large at {${\Dl\simeq20~}$Mpc}. {The model accurately reconstructs $\fpeak$ at approximately 18~Mpc. However, a detection of an event at this distance may be considered an optimistic scenario (see~\cite{Sachdev_2020}).}

The $\apr{20}$ model successfully reconstructs the injected signal. However, it reaches lower median $\Ov$ values than the other models. At low luminosity distances, the $\apr{20}$ model achieves {a median $\Ov$ of ${0.964}$}. This trend is in agreement with the results discussed in Sec.~\ref{Performance of KNN models}, i.e., the mean $\F$ for the $\apr{20}$ model is lower than those of the other models, $\ntrain\geq 40$, as depicted in Fig.~\ref{fig:mean_std_plots.pdf}.

The injection study provides additional information on the reconstructive capacity of the KNN models under realistic conditions. We observe very good agreement between the models with $\ntrain \geq 40$, and thus conclude that a training set of $40$ simulations suffices for the construction of robust templates.

\subsection{Dependence on EOS}\label{EOS dependence}

As discussed in Sec.~\ref{results}, a training set with a minimum size of $\ntrain=40$ suffices to build a robust model that leads to large values of {$\F$} and successful injection studies {for a fixed EOS model and equal-mass binaries}. {To further test our findings, we repeat our analysis for the construction of a KNN model using }the SFHX EOS.

{We only consider the case where $\ntrain=40$, since we previously determined that such a size is sufficient to achieve convergence.} As before, we split the data set into training ($\ntrain=40$), validation ($\nval=29$), and test ($\ntest=30$) sets. The validation sets between the APR4 and SFHX KNN models have different sizes since the total number of simulations for each EOS model is different. {Despite these differences and inconsistencies, the comparison is still meaningful {as it enables us to} examine the robustness of the predictions{, for models constructed with KNN regression,} in terms of unknown GW signals, from a test set of simulations, covering a significant binary mass range. In the following, we denote the KNN models with training set of size $\ntrain=40$ for the SFHX and APR4 EOS models as $\apr{40}$ and $\sfhx{40}$, respectively. }

\begin{figure}[h!]
	\includegraphics[scale=0.36]{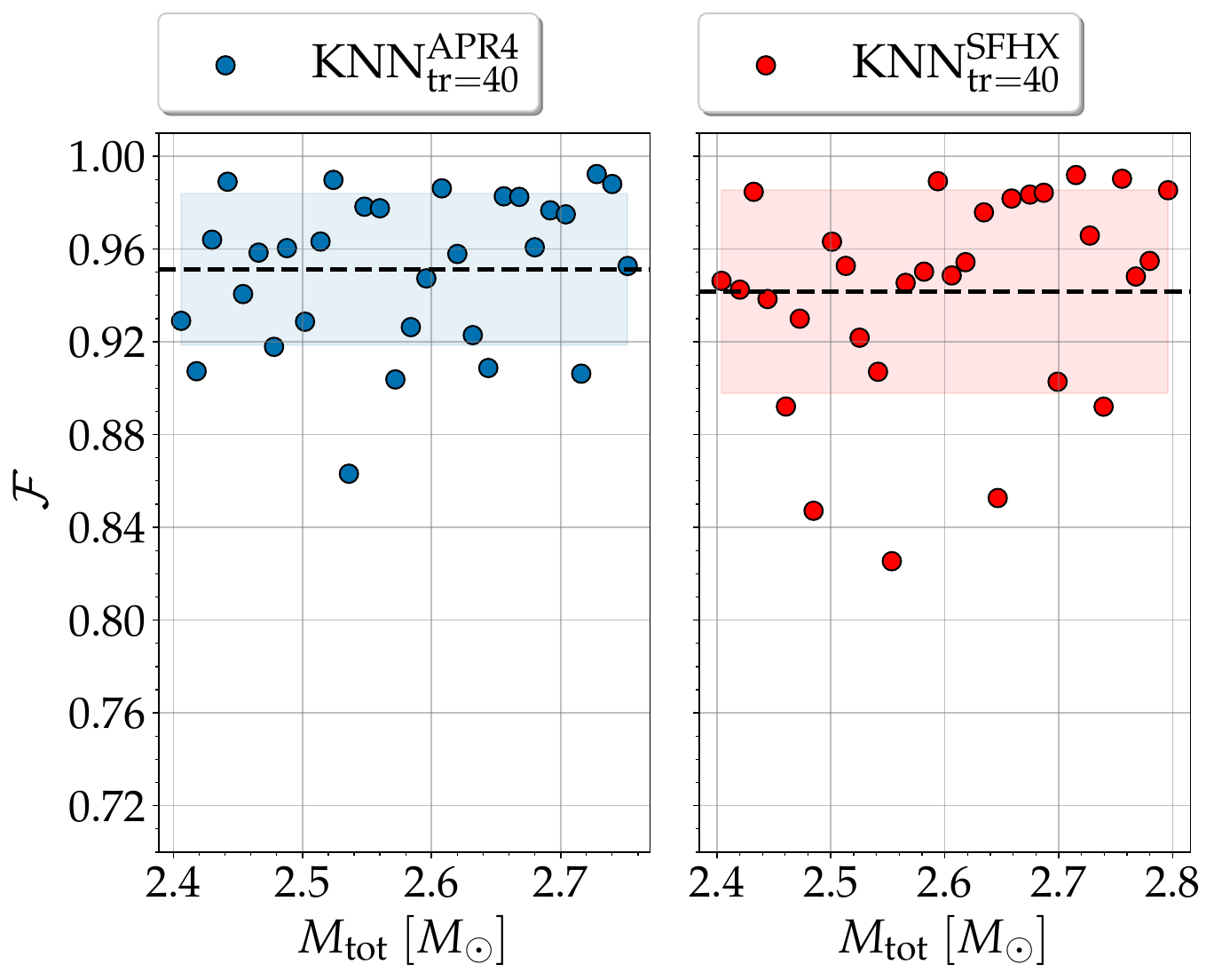}
	\caption{\label{fig:COMPARISON_test_FFs.pdf} Faithfulness $\F$ as a function of $\mtot$ for the  $\apr{40}$ (blue) and  $\sfhx{40}$ (red) KNN models constructed using a training set of size $\ntrain=40$.}
\end{figure}

Figure~\ref{fig:COMPARISON_test_FFs.pdf} shows the $\F$ values as a function of the total binary mass $\mtot$ for the unseen GW signals of the test sets of the APR4 and SFHX EOS models. It is evident from Fig.~\ref{fig:COMPARISON_test_FFs.pdf} that the general behavior of the data is qualitatively similar in both cases. Both KNN models exhibit high values of {$\F$} where for most of the GW signals the {$\F$} values range roughly between 0.90 and 1.0, with some outliers that still have $\F>0.80$. {The mean value of $\F$ for the $\sfhx{40}$ model is slightly lower ($\simeq 1\%$ difference)}. Both values of {$\F$} are high, considering that these signals are not used for training or validation. Furthermore, the $\sfhx{40}$ predictions with the highest {$\F$} reach values close to $1$ (perfect match) as observed for $\apr{40}$. The $\sfhx{40}$ model exhibits slightly larger scatter (and more outliers) compared to the $\apr{40}$ case. However, the lowest values of {$\F$} for the $\apr{40}$ and $\sfhx{40}$ models exhibit comparable values, i.e., larger than $0.85$ and $0.82$, respectively. 

The good agreement between the two KNN models suggests that our results are robust {in terms of the {$\F$} values achieved}, and in fact a training set with a minimum size of $40$ simulations may be sufficient to construct accurate templates. We also note that, for practical use, the size $\ntrain$ should be converted to number density defined as 
\begin{eqnarray}
n&=&\frac{\ntrain}{\mtot^\mathrm{max}-\mtot^\mathrm{min}}\,,
\end{eqnarray}
where $\mtot^\mathrm{max}$ ($\mtot^\mathrm{min}$) refers to the maximum (minimum) total binary mass of the respective data set. Table~\ref{KNN number densities} reports the number densities and the corresponding number of simulations per $0.1~M_\odot$, for all KNN models used in this work. For both EOS models, we considered comparable $\mtot$ ranges, and thus, the choice of $\ntrain$ corresponds roughly to the same number density. For a stiffer EOS model, where binary configurations with higher $\mtot$ can occur, the corresponding $\ntrain$ value may need to be adjusted.

\begin{table} 
	\begin{ruledtabular}
		\begin{tabular}{c|cccccc}
                \textrm{EOS}&
                \textrm{$\ntrain$}&
                \textrm{$n~[\frac{\mathrm{sims.}}{M_\odot}]$}&
   \textrm{$N_\mathrm{per~0.1M_\odot}$}&
   \textrm{$[\mtot^\mathrm{min},\mtot^\mathrm{max}]$}&
   \textrm{{$\mathrm{mean}~\F$}}\\
    \hline

APR4 & 20 & 55.87 & 6 & 2.40, 2.76& 0.938\\
APR4 & 40 & 111.73 & 11 & 2.40, 2.76& 0.951\\
APR4 & 60 & 167.60 & 17 & 2.40, 2.76& 0.953\\
APR4 & 80 & 223.46 & 22 & 2.40, 2.76& 0.953\\
APR4 & 100 & 279.33 & 28 & 2.40, 2.76& 0.956\\
\hline
SFHX & 40 & 100.00 & 10 & 2.40, 2.80& 0.941\\
		\end{tabular}
	\end{ruledtabular}
	\caption{Number densities and number of simulations per 0.1$M_\odot$ for all KNN models considered in this work.}
	\label{KNN number densities}
\end{table}

\begin{figure}[h!]
	\includegraphics[scale=0.34]{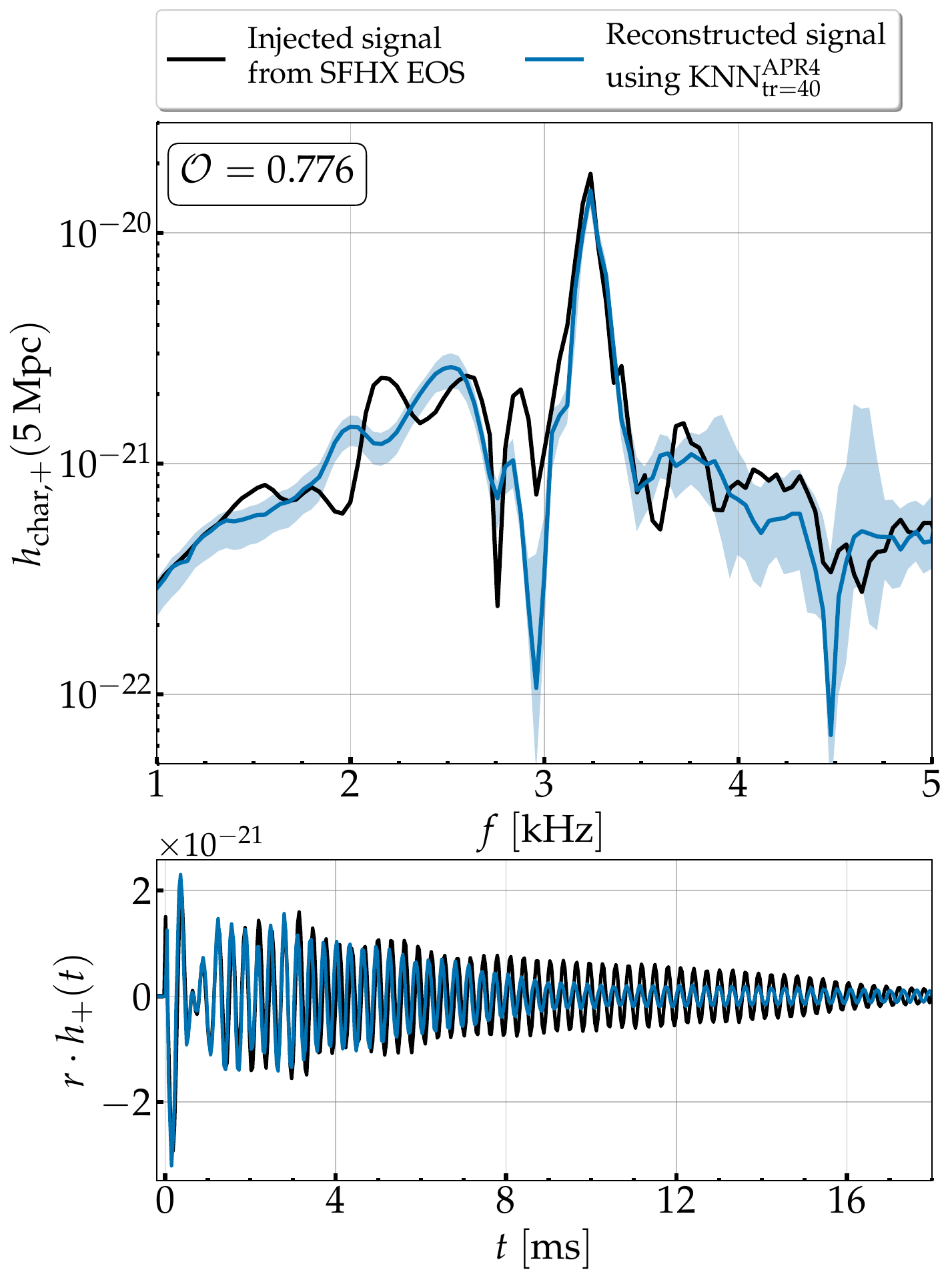}
	\caption{\label{fig:injection_of_sfhx_with_apr4.pdf} Reconstructed signal (blue) using the $\apr{40}$ model for the injected signal (black) of the simulation from the SFHX test set. {Blue shaded areas represent the 90\% credible intervals of the posterior probability of the GW spectra.}}
\end{figure}

{We conduct an additional test to determine whether the KNN models discussed in this work, constructed for single EOSs, can be used to reconstruct general signals, i.e., those originating from simulations where other EOSs are employed.} {We inject a signal from the SFHX test set, with $\mtot=2.796~M_\odot$, and use the $\apr{40}$ as reconstruction model. This configuration is chosen so that the characteristic frequencies of its GW spectrum are in the frequency regime of the $\apr{40}$ model. Both EOS models represent relatively soft EOSs, so there is overlap in the frequency range of their respective $\fpeak$ as a function of mass. {For a value of $\mtot$ that corresponds to an $\fpeak$ that is not within the characteristic frequency range of the GW spectra of the $\apr{40}$ model, a reconstruction with relatively high $\Ov$ is not possible.} We consider a luminosity distance of $\Dl=5~$Mpc and sample $\mtot, \phi_0, t_0, \Dl$ using the setup described in Sec.~\ref{Injections}.} We find that the $\apr{40}$ model is capable of reconstructing the injected SFHX signal. Figure~\ref{fig:injection_of_sfhx_with_apr4.pdf} displays the GW spectra and time-domain signals for the injected signal {and the reconstructed signal, which maximizes the likelihood}}. The agreement is impressive, achieving a median overlap of {$\Ov=0.776$}. The spectra match well, particularly for the value of $\fpeak$, and there is also relatively good agreement in the subdominant feature, at roughly 2.5~kHz. Therefore, $\apr{40}$ manages to capture the dominant and approximately the subdominant frequency peaks. 

The injected luminosity distance of the source is not reconstructed accurately. This is because the injected signal corresponds to a system with $\mtot=2.796~M_\odot$, while the total mass for the $\apr{40}$ model that best reconstructs the injected signal is $\mtot=2.41~M_\odot$, which is a much lighter binary, resulting in weaker excitation of the quadrupolar mode. {Therefore, these models, in their current state, may not be used to reconstruct the total mass of the binary system.}

{The observation that a model constructed for one EoS can adequately reconstruct signals of another EOS may indicate that a fully EOS-dependent post-merger model, {i.e., a model that is suitable for any EOS,} could potentially be constructed with simulations for a large but possibly manageable number of EOSs.} {In addition, it may be possible to introduce modifications to extend the capabilities of the models we presented in this work.} For example, {shifts in the frequency-domain may be employed to ensure that KNN models, constructed for a single EOS, cover a wider range of $\fpeak$ values that would allow {approximate} reconstruction of signals from stiffer {or softer} EOS models than those discussed here. }


\section{Conclusions}\label{Conclusions}

In this work, we construct {accurate} models for the post-merger GW emission from BNS merger remnants. We use a supervised learning method, i.e., KNN regression, and create {a large} library of numerical relativity simulations, for a single EOS model (APR4), which includes 157 equal-mass binaries. We split the entire data set into subsets used for training, validation, and testing. To determine the optimal size of the data set required for the construction of reliable templates, we vary the size of the training set for values in $\ntrain =\{ 20, 40, 60, 80, 100\}$ and optimize the corresponding KNN models. 

We evaluate the performance of the KNN models in terms of the distribution of the faithfulness $\F$ values for the simulations of the test set. The test set is a holdout set of 30 simulations covering the entire $\mtot$ range that enables us to compare the various KNN {models}. We further explore the robustness of the KNN models by performing injection studies using the three-detector network of Advanced LIGO and Advanced Virgo with a varied luminosity distance of the source. We find that the KNN models achieve {high values of }$\F$ for completely unseen GW signals from the test set. The predictions with the highest {$\F$}, reach values ranging from $0.980$ to $0.995$, which is remarkably high, considering these signals are not used during the optimization of the KNN models. We find that KNN models with $\ntrain \geq 40$, exhibit convergence with mean $\F$ approximately equal to $0.95$, {and standard deviation approximately around $0.032$,} for the simulations of the holdout test set (see Fig.~\ref{fig:mean_std_plots.pdf}).

{Furthermore, all KNN models successfully reconstruct the injected signals.} Using the posterior probabilities of the $\mtot, \phi_0$, $t_0$, and $\Dl$, we derive the distributions of $\fpeak$ and {$\Ov$}, and show that the ability to reconstruct $\fpeak$ scales with the luminosity distance, as expected. Models with $\ntrain\geq 40$ exhibit convergence as the medians of $\fpeak$ approach the injected value. The median $\Ov$ values, for models with $\ntrain\geq 40$, are in the range of $0.968$ to $0.992$, while the $\apr{20}$ model achieves lower values.

{All KNN models, for} sampling rates comparable to those employed by Advanced LIGO-Virgo, {exhibit} execution times {that} range from $1$ to $3~\mathrm{ms}$, and only slightly increase for unrealistically large sampling rates. Thus, such models can indeed be employed in real detection searches{, that require million of waveform evaluations,} with future detectors.

Furthermore, {we verify our findings} using an additional sequence of equal-mass binaries for the SFHX EOS. For this data set, the mean {$\F$} {and standard deviation} for a holdout test set of 30 simulations is $0.941$ {and $0.0437$, respectively}. This large value {of mean $\F$} qualitatively agrees with the values for the KNN models of the APR4 data set. This further supports the argument that a size of $\ntrain=40$, which corresponds to a similar number density of simulations per unit of mass for these two EOS models, suffices for the construction of accurate GW templates.

We show that the APR4 KNN models may potentially be used to reconstruct the dominant component of GW signals from different EOS models. To accomplish this, we perform an injection with a signal from the SFHX data set and utilize the $\apr{40}$ model as a reconstruction tool. We find that the $\apr{40}$ model, provided a sufficiently large $\snr$, accurately reconstructs the $\fpeak$ component of the GW spectrum and achieves an {overlap} of $\Ov=0.776$. {This result indicates that the impact of the EOS might be less significant than expected, and thus the construction of post-merger models {that cover all possible EOSs}, may be accomplished using a moderate number of simulations.} {However, it is clear that this test is limited to only a particular region in the parameter space of possible EOS models. The GW spectra of these binary configurations exhibit similarities that may not be present for other EOS models.}

Finally, we conclude that a minimum size of $\ntrain=40$ simulations used for training, or equivalently a number density of approximately $11$ simulations per $0.1~M_\odot$ of total binary mass, is sufficient {(mean $\F\simeq 0.95$)} for the construction of reliable and accurate GW models {for fixed EOS and binary mass ratio of $q=1$.} {For models with training sets of $\ntrain\geq40$, the mean values of $\F$ for the simulations of the test set converge to the value $\simeq 0.95$. In those models, the maximum median $\Ov$ values are in the range of $0.969$ to $0.992$, while the $\apr{20}$ model achieves lower $\Ov$ values.}

In this work, we have addressed the first of a series of questions toward the construction of a general purpose EOS-dependent GW model for the post-merger emission {based on supervised learning methods}.  We systematically investigated the minimum size of the data set required for developing accurate templates, {using the supervised learning method of KNN regression}, for a sequence of equal-mass binaries and a fixed EOS model. {Our findings are based on this particular method, however, the KNN regression may be indicative of other types of regression methods.} However, it is evident that the models discussed in this work need to be extended in order to be applicable to realistic scenarios where the EOS is only partially known. {The intrinsic parameters have to be adjusted to include the effects of the EOS and binary mass ratio}. In this regard, varying the additional degrees of freedom would significantly increase the number of simulations needed, as discussed in~\cite{2022arXiv220106461W}. {A potential caveat of any model based on simulations is the fact that the simulated GW signal depends on the employed resolution and may be affected by missing physics in the simulations or the finite number of EOS models.} {However, the first few milliseconds of the post-merger phase, in which most of the GW radiation occurs, may be more reliable than the later evolution, i.e., of the order of tens of milliseconds.} {In addition, the fact that an APR model successfully recovers a SFHX signal may indicate that these issues are less relevant in practice.}

In future work, we plan to investigate the behavior of the remaining degrees of freedom, such as the EOS model, or the binary mass ratio $q$, and construct templates for the post-merger GW emission that consider those quantities as intrinsic parameters. {The effect of the binary mass ratio is likely {less dominant} in comparison to that of the EOS model. This is because the GW spectra for unequal-mass configurations exhibit similarities to those of equal-mass systems, while the {range of} characteristic frequencies of the GW spectra are not expected to differ significantly. {However, the mechanisms that explain some of the subdominant frequency peaks, such as $\fspiral$ (see~\cite{AB2015}), in the GW spectrum depend on the mass ratio, and thus the amplitudes of those may change for binaries with low values of~$q$.} The EOS model affects the dynamics of the post-merger system, and thus the {characteristic frequencies of the} GW spectra. {To obtain faithful templates, that achieve large $\Ov$ values, the model must accurately describe the the dominant and subdominant frequency components of the GW spectrum.} For this reason, the model must incorporate EOSs, with different degrees of softness and stiffness. }

The data underlying this article will be shared on reasonable request to the corresponding author. {The KNN models are available in~\cite{GSI_repo}. }

\begin{acknowledgments}
T.S. thanks Georgios Lioutas, and Christian Schwebler for the useful discussions. T.S. and A.B. acknowledge support by the State of Hesse within the Cluster Project ELEMENTS. A.B. acknowledges support by Deutsche Forschungsgemeinschaft (DFG, German Research Foundation) through Project No. 279384907—SFB 1245 (subproject B07) and by the European Research Council (ERC) under the European Union’s Horizon 2020 research and innovation program under ERC Grant HEAVYMETAL No. 101071865. K.C. acknowledges support from the Department of Energy under award number DE-SC0023101 and the Sloan Foundation. {The work of K.M.\ and F.K.R.\ is supported by the Klaus Tschira Foundation. F.K.R. acknowledges funding by the European Union (ERC, ExCEED, project number 101096243). Views and opinions expressed are however those of the author(s) only and do not necessarily reflect those of the European Union or the European Research Council Executive Agency. Neither the European Union nor the granting authority can be held responsible for them. N.S. acknowledges support by Virgo, which is funded, through the European Gravitational Observatory (EGO), by the French Centre National de Recherche Scientifique (CNRS), the Italian Istituto Nazionale di Fisica Nucleare (INFN) and the Dutch Nikhef, with contributions by institutions from Belgium, Germany, Greece, Hungary, Ireland, Japan, Monaco, Poland, Portugal, Spain.}
\end{acknowledgments}

\appendix
\section{KNN models}\label{KNN models}

{In this section, we provide additional information regarding the KNN models. We show histograms of the mass distribution for simulated binary configurations. We present the mean and standard deviation of {$\F$} for the simulations of the test set for the various training set sizes employed in this work. Furthermore, we discuss the {$\F$} values of the simulations of the validation sets.}

\subsection{Data set of simulations}
Figure~\ref{fig:test_sets_complete_set.pdf} shows the histograms of the simulated binaries for the APR4 and SFHX EOS models. The complete set consists of simulations uniformly distributed with respect to total binary mass. The test set includes simulations distributed in an equidistant manner. 

\begin{figure}[h!]    
	\includegraphics[scale=0.38
]{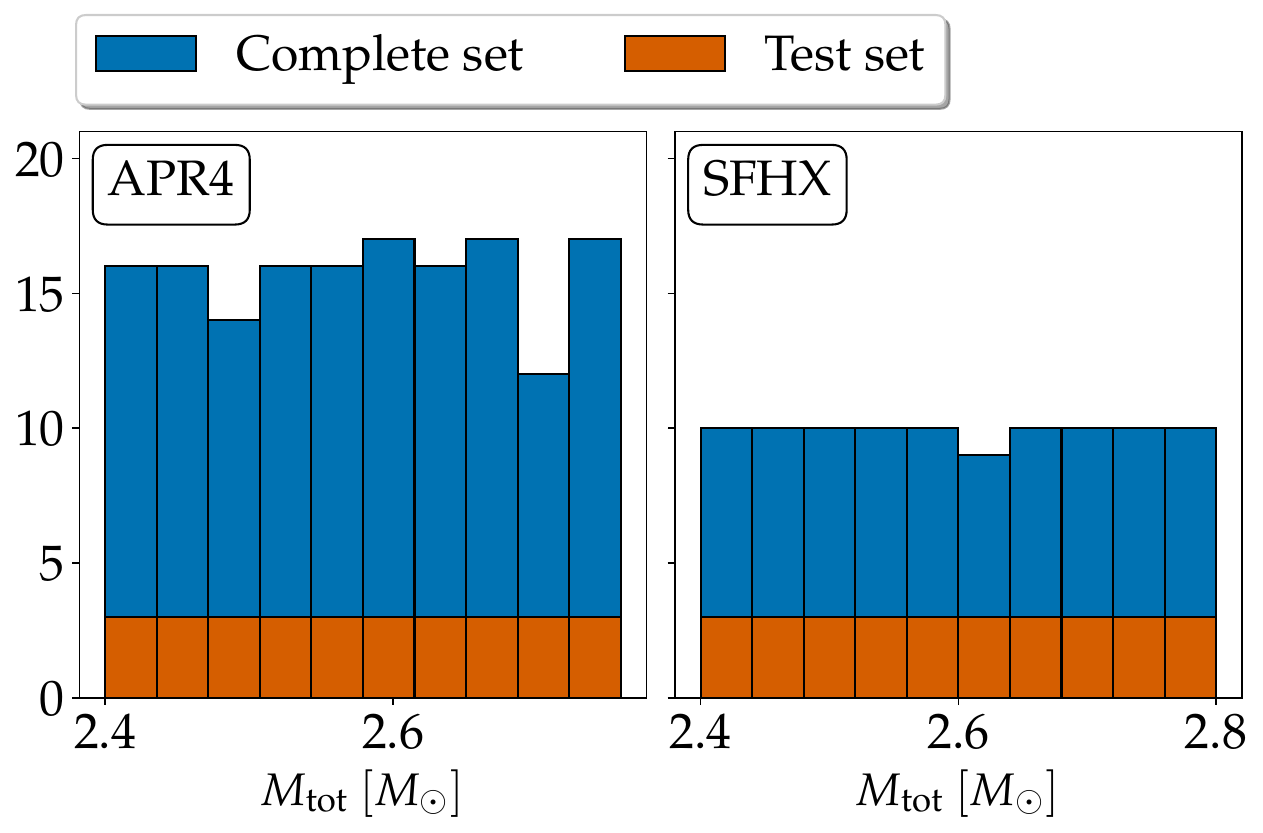}
\caption{\label{fig:test_sets_complete_set.pdf} Histograms for the simulated binaries for the APR4 (left) and SFHX (right) EOS models. Blue histograms indicate the simulations of the complete set. Orange histograms indicate the configurations of the test set.}
\end{figure}

\subsection[Validation data]{Validation data}\label{Validation data}

\begin{figure}[h]    
	\includegraphics[scale=0.36]{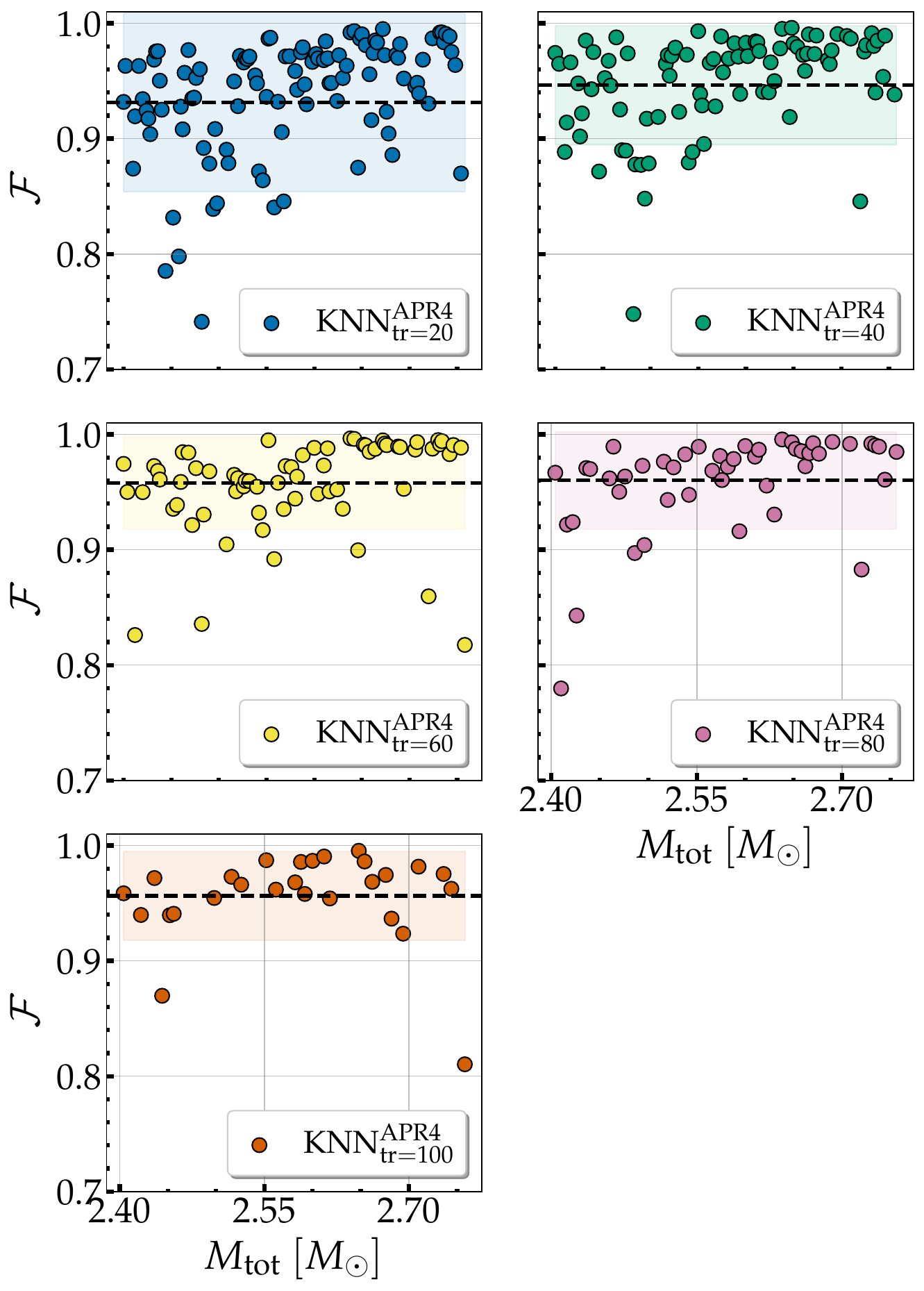}
	\caption{\label{fig:val_FFs.pdf} {$\Fs$} for the signals of the validation set, which is used for hyper-parameter tuning, as a function of total binary mass $\mtot$ for the all KNN models.}
\end{figure}

The validation sets have different sizes (for the various KNN models) that are supplementary to the training sets. The validation sets are used during the grid search for the hyper-parameter tuning to determine the optimal {hyper-}parameters of the KNN models. Figure~\ref{fig:val_FFs.pdf} shows the {$\F$} as a function of the total binary mass $\mtot$ for the configurations of the validation sets. As in the case of the test set comparison, we find that the mean value of the {$\F$} {exhibits an increase as the size of the training set increases}. For the $\apr{20}$ model the mean value slightly drops, however, it still corresponds to a large value of {0.931}. The scatter of the data around the mean, expressed by the standard deviation, also decreases as the size $\ntrain$ of the training set increases, and thus the resolution of the KNN model increases.

In addition, all KNN models exhibit some outliers that decrease as $\ntrain$ increases. These outliers tend to occur at the corners of the parameter space (low-mass and high-mass configurations). {This is a well-known characteristic of the KNN schemes as these methods rely on neighbors to make accurate predictions, and thus configurations at the edge of the parameter space have more one-sided neighbors.} This effect is possibly minimized in the test data, as we did not include the two simulations at the edges (minimum and maximum $\mtot$) and also used a uniform distribution of configurations covering the entire range. {We also observe a slight preference for outliers to occur at low $\mtot$, especially for the $\apr{20}$ model.}  
\begin{figure}[h!]    
	\includegraphics[scale=0.38]{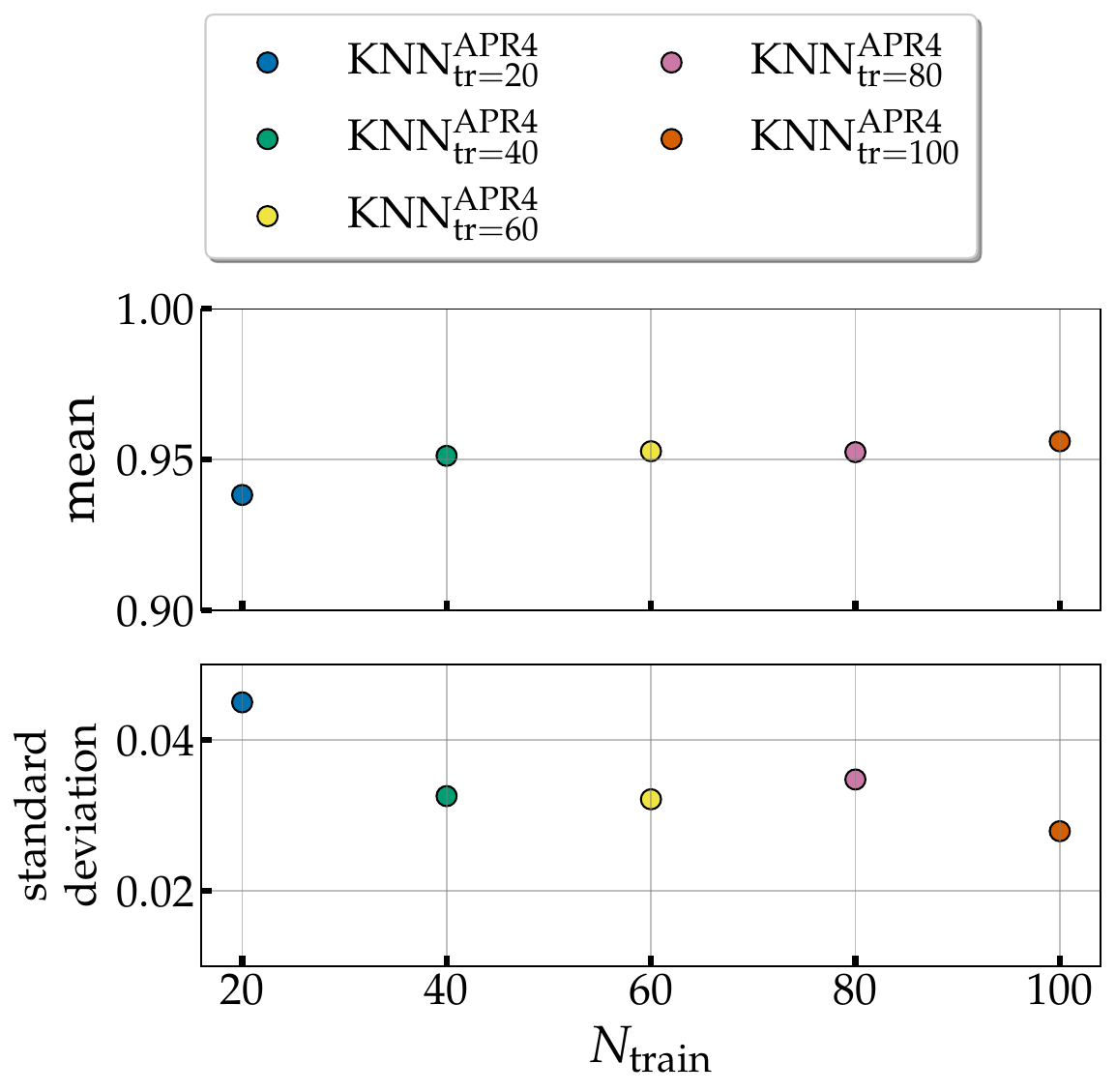}
	\caption{\label{fig:mean_std_plots.pdf} Upper panel: Mean {$\F$} as a function of the size of the training set $\ntrain$. Bottom panel:  Standard deviation as a function of the size of the training set $\ntrain$. }
\end{figure}

\subsection{Test data}
{Figure~\ref{fig:mean_std_plots.pdf} displays the mean and standard deviation for the KNN models. There is a small systematic increase in the mean {$\F$} as $\ntrain$ increases. Furthermore, the standard deviation decreases as $\ntrain$ increases. }

\section{{Parameter estimation with next-generation detectors}}
{In this section, we present the posterior probabilities of $\fpeak$ and $\Ov$ for simulated detections using the next-generation detector, the Einstein Telescope~\cite{Maggiore:2019uih}. To achieve this, we first compute the ratio $X$ of the $\snr$ for Advanced LIGO and Advanced Virgo over the $\snr$ for the Einstein Telescope. Then using $X$ and fixed values of $\snr$, we derive the corresponding luminosity distance for simulated detections compatible with the sensitivity curve of the Einstein Telescope. Figure~\ref{fig:injections_fpeak_recon_and_FF_ET.pdf} depicts the {90\% credible interval for} $\fpeak$ (left) and the corresponding $\Ov$ (right) as a function of the luminosity distance of the injected signal for the various KNN models.  }

\begin{figure*}[t!]    
	\includegraphics[scale=0.345]{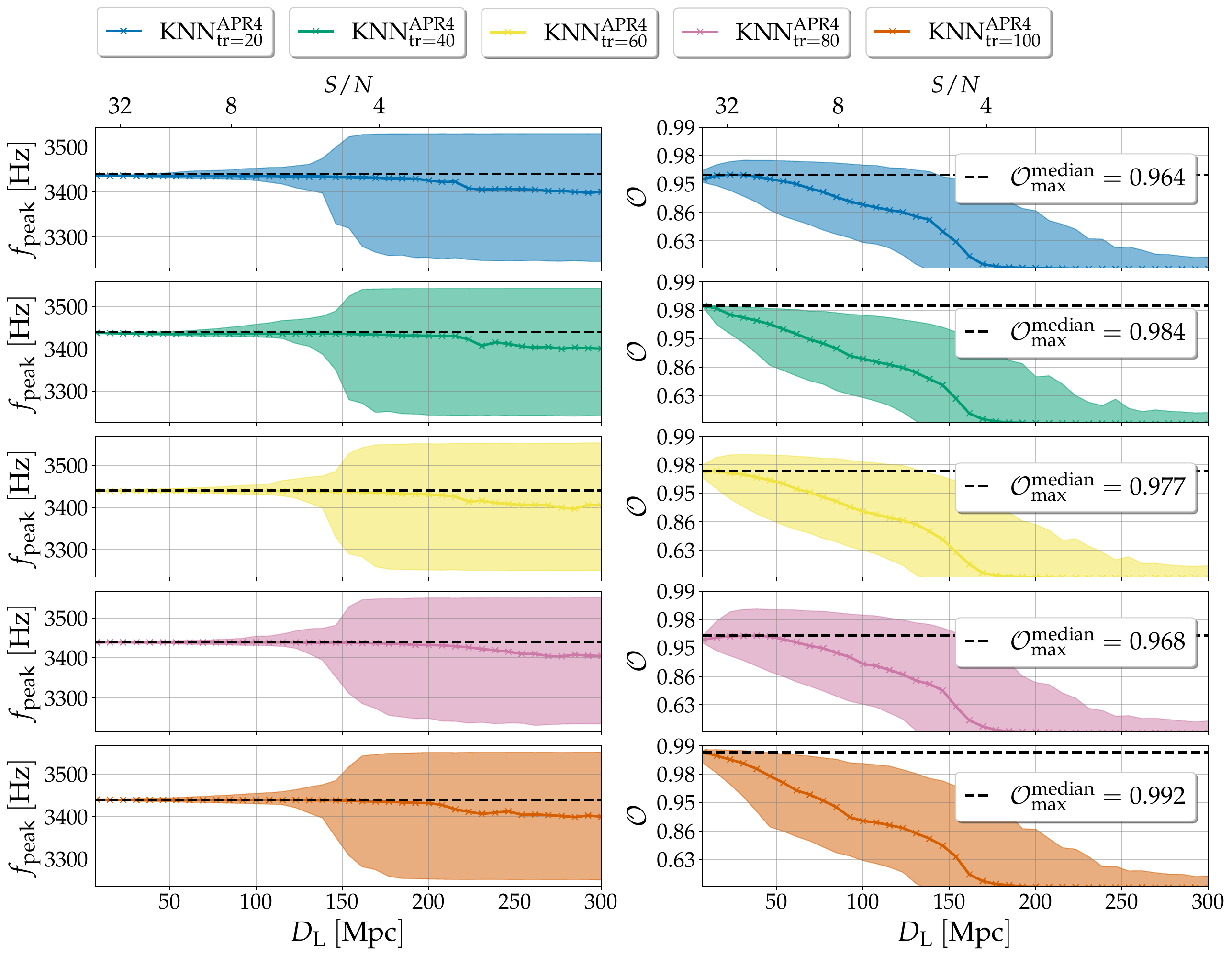}
\caption{\label{fig:injections_fpeak_recon_and_FF_ET.pdf}{Same as in Fig.~\ref{fig:injections_fpeak_recon_and_FF.pdf} but for the sensitivity curve of the Einstein Telescope~\cite{Maggiore:2019uih}.}}
\end{figure*}

\section{Best and worst predictions for the test data}\label{Best predictions for the unseen signals}

{In this section, we present the predictions of the $\apr{40}$ and $\apr{100}$ KNN models for the configurations of the test set that lead to the highest and lowest values of {$\F$}. Figures~\ref{fig:dumpGW_model_best_worst_040.pdf} and \ref{fig:dumpGW_model_best_worst_100} show the predicted GW signals in the time domain in comparison to the corresponding simulated signals. This comparison enables us to better understand how {$\F$}, which is a global metric for the match between two signals, is affected by the different segments of the signal. For example, fits with high {$\F$} show a good match over the whole duration of the signal. In contrast, fits with low {$\F$} exhibit {certain periods within the signal} where there is an apparent mismatch between the oscillations that reduce the overall faithfulness {$\F$}.}

\begin{figure*}[h]    
	\includegraphics[scale=0.27]{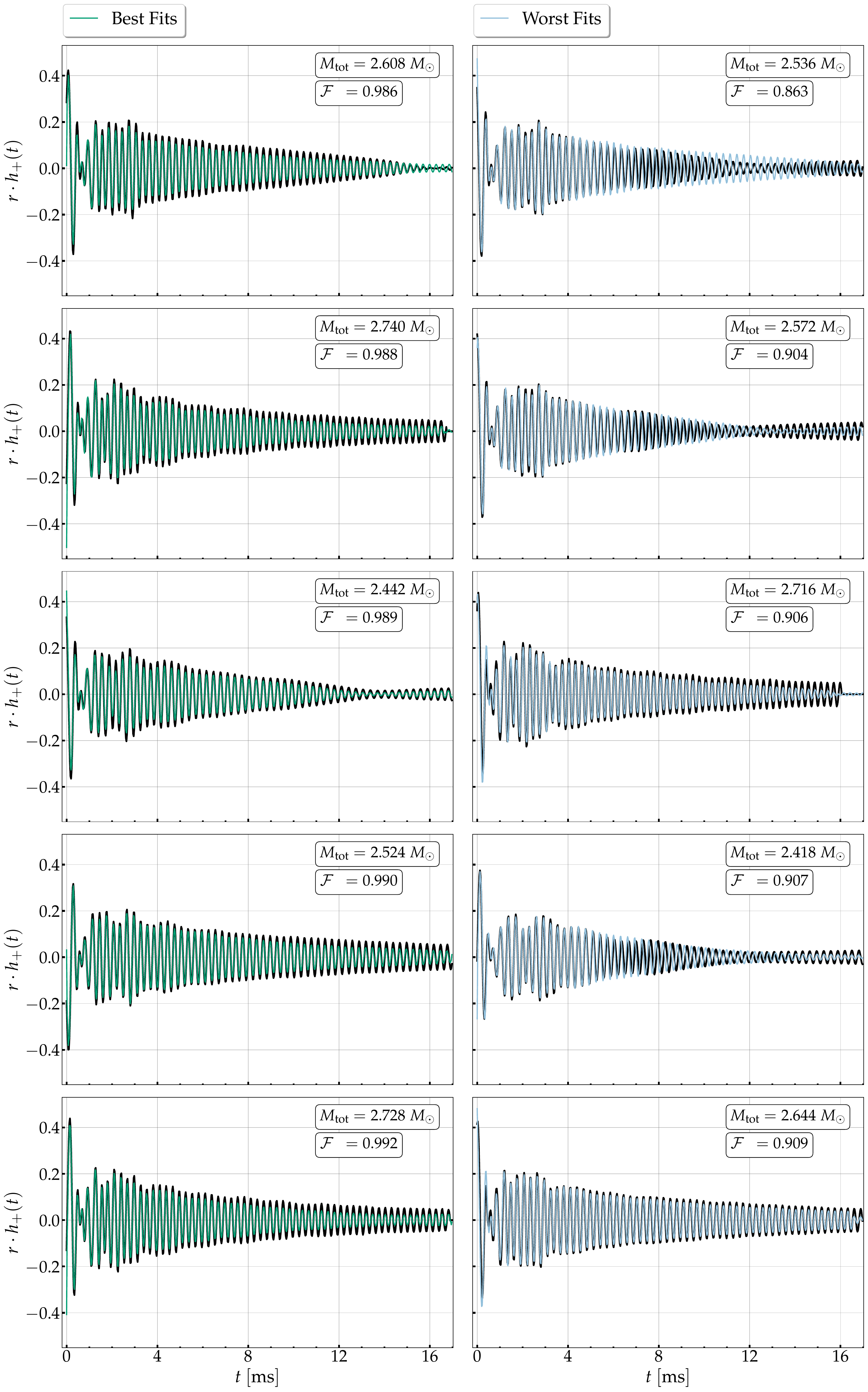}
\caption{\label{fig:dumpGW_model_best_worst_040.pdf} Comparison between the predictions of the $\apr{40}$ model that lead to the highest (green) and lowest (cyan) {$\F$} values and the simulated GW signals (black) for the binary configurations of the test set. }

\end{figure*}
\begin{figure*}[h]    
	\includegraphics[scale=0.27]{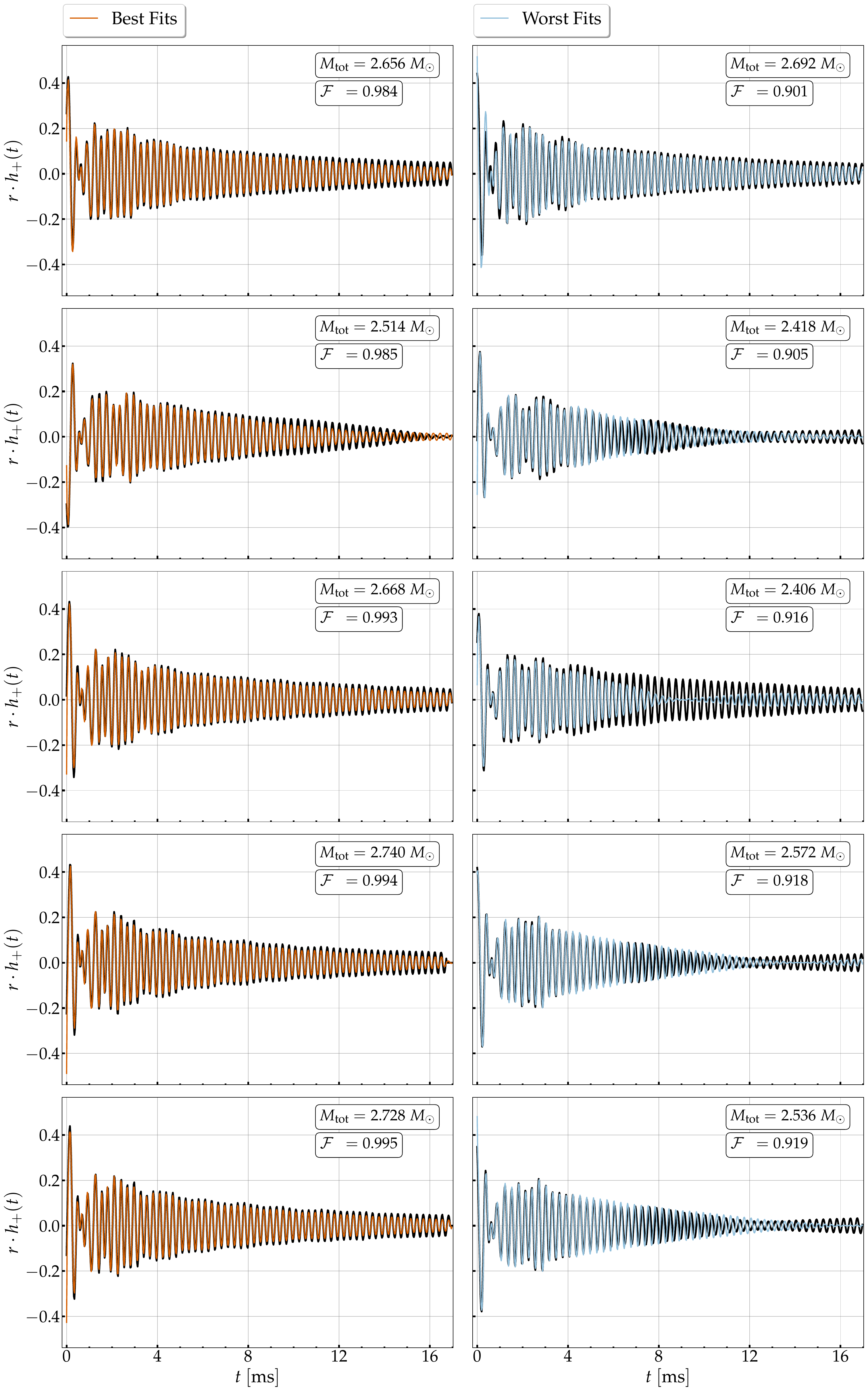}
\caption{\label{fig:dumpGW_model_best_worst_100}Comparison between the predictions of the $\apr{100}$ model that lead to the highest (orange) and lowest (cyan) {$\F$} values and the simulated GW signals (black) for the binary configurations of the test set. }
\end{figure*}


%

\end{document}